\documentclass[twocolumn]{aastex7}

\usepackage{amsmath}
\usepackage[figuresright]{rotating}

\begin{document}
  
\title{\texttt{LEO-Vetter}: Fully Automated Flux- and Pixel-Level Vetting of TESS Planet Candidates to Support Occurrence Rates}

\author[0000-0001-9269-8060]{Michelle Kunimoto}
\email{mkuni@phas.ubc.ca}
\correspondingauthor{mkuni@phas.ubc.ca}
\affiliation{Department of Physics and Astronomy, University of British Columbia, 6224 Agricultural Road, Vancouver, BC V6T 1Z1, Canada}
\affiliation{Department of Physics and Kavli Institute for Astrophysics and Space Research,\\ Massachusetts Institute of Technology, Cambridge, MA 02139, USA}

\author[0000-0003-0081-1797]{Steve Bryson}
\affiliation{NASA Ames Research Center, Moffett Field, CA 94035, USA}
\email{steve.bryson@nasa.gov}

\author[0000-0001-6513-1659]{Drayson Jaffee}
\affiliation{Marin Academy, San Rafael, CA 94901, USA}
\email{dijaffee@gmail.com}

\author[0000-0002-5904-1865]{Jason F. Rowe}
\affiliation{Department of Physics \& Astronomy, Bishop's University, 2600 Rue College, Sherbrooke, QC J1M 1Z7, Canada}
\email{jrowe@ubishops.ca}

\author[0000-0002-6939-9211]{Tansu Daylan}
\affiliation{Department of Physics and McDonnell Center for the Space Sciences, Washington University, St. Louis, MO 63130, USA}
\email{tansu@wustl.edu}

\author[0000-0002-8965-3969]{Steven Giacalone}
\altaffiliation{NSF Astronomy and Astrophysics Postdoctoral Fellow}
\affiliation{Department of Astronomy, California Institute of Technology, Pasadena, CA 91125, USA}
\email{giacalone@astro.caltech.edu}

\author[0000-0001-6513-1659]{Jack J. Lissauer}
\affiliation{NASA Ames Research Center, Moffett Field, CA 94035, USA}
\email{jack.lissauer@nasa.gov}

\author[0000-0002-1119-7473]{Michael R. B. Matesic}
\affiliation{D{\'e}partement de Physique, Trottier Institute for Research on Exoplanets, Ciela Institute for Computation \& Astrophysical Data Analysis, Universit{\'e} de Montr{\'e}al, 1375 Th{\'e}r{\`e}se-Lavoie-Roux Av., Montr{\'e}al, QC H2V 0B3, Canada}
\affiliation{Department of Physics \& Astronomy, Bishop's University, 2600 Rue College, Sherbrooke, QC J1M 1Z7, Canada}
\email{mrbmatesic@uwaterloo.ca}

\author[0000-0001-7106-4683]{Susan E. Mullally}
\affiliation{Space Telescope Science Institute, 3700 San Martin Drive, Baltimore, MD 21218, USA}
\email{smullally@stsci.edu}

\author[0009-0006-6397-2503]{Yoshi Nike Emilia Eschen}
\affiliation{Department of Physics, University of Warwick, Gibbet Hill Road, Coventry, CV4 7AL, UK}
\affiliation{Department of Physics and Kavli Institute for Astrophysics and Space Research,\\ Massachusetts Institute of Technology, Cambridge, MA 02139, USA}
\email{yoshi.eschen@warwick.ac.uk}

\begin{abstract}
The Transiting Exoplanet Survey Satellite (TESS) has identified several thousand planet candidates orbiting a wide variety of stars, and has provided an exciting opportunity for demographic studies. However, current TESS planet searches require significant manual inspection efforts to identify planets among the enormous number of detected transit-like signatures, which limits the scope of such searches. Demographic studies also require a detailed understanding of the relationship between observed and true exoplanet populations; a task for which current TESS planet catalogs are rendered unsuitable by the subjectivity of vetting by eye. We present \texttt{LEO-Vetter}, a publicly available and fully automated exoplanet vetting system designed after the Kepler Robovetter, which is capable of efficiently producing catalogs of promising planet candidates and making statistically robust TESS demographic studies possible. \texttt{LEO-Vetter} implements flux- and pixel-level tests against noise/systematic false positives and astrophysical false positives. The vetter achieves high completeness (91\%) and high reliability against noise/systematic false alarms (97\%) based on its performance on simulated data. We demonstrate the usefulness of the vetter by searching $\sim200,000$ M dwarf light curves, and reducing $\sim20,000$ transit-like detections down to 172 uniformly vetted planet candidates. \texttt{LEO-Vetter} facilitates analyses that would otherwise be impractical to perform on all possible signals due to time constraints or computational limitations. Users will be able to efficiently produce their own TESS planet catalog starting with transit-like detections, as well as have the framework needed to characterize their catalog’s completeness and reliability for occurrence rates.
\end{abstract}

\keywords{}

\section{Introduction}

The search for exoplanets has been revolutionized by space-based telescopes like Kepler \citep{Borucki2010,Lissauer2023} and the Transiting Exoplanet Survey Satellite \cite[TESS;][]{Ricker2015}. These missions have generated an unprecedented volume of data, leading to the identification of thousands of potential transiting planets. Kepler, during its primary mission, monitored over 150,000 stars and detected more than 4,000 planets and planet candidates \citep{Thompson2018}. TESS, designed to survey tens of millions of stars across the entire sky, has discovered more than 6000 planet candidates since its launch in 2018 \citep{Guerrero2021}.\footnote{Based on the TESS Object of Interest Catalog on ExoFOP, accessed 2025 April 08 \citep{ExoFOP}}

The sheer volume of data from these missions presents both an opportunity and a challenge. The traditional methods of manually vetting planet candidates are time-consuming and labor-intensive, making it impractical to keep up with the influx of new data. Human inspection, while thorough, is also inherently subject to biases such as pattern recognition errors, fatigue, and subjective judgment.

The need for automated and efficient vetting processes has become increasingly apparent. Automated vetting systems can rapidly analyze light curves and assess the likelihood of potential signals of being true planetary transits versus false positives caused by astrophysical or instrumental noise. Fully automated vetting algorithms also produce uniform, reproducible planet catalogs suitable for studies of planetary demographics. Statistically robust demographics calculations require a detailed understanding of the relationship between observed and true exoplanet populations; a task for which manually-produced planet catalogs are rendered unsuitable by the subjectivity of vetting by eye.

Numerous vetting efforts emerged to classify transit signals detected in Kepler data automatically and uniformly. The Q1-Q17 DR24 planet catalog was the first Kepler catalog resulting from fully uniform vetting \citep{Coughlin2016}, with possible transit signals (Threshold Crossing Events, or TCEs) dispositioned as either planet candidates or false positives using the Kepler ``Robovetter.'' The Robovetter was a classical decision tree classifier with criteria designed to mimic the manual process for rejecting false positives. The Kepler team also developed a machine learning-based classifier known as the ``Autovetter'' to produce an ancillary DR24 planet catalog \citep{Catanzarite2015, McCauliff2015}. The final Kepler data release, generally referred to as DR25 \cite[][but hereafter denoted KDR25 to emphasize that it is a product of the Kepler mission rather than TESS]{Thompson2018} again adopted the Robovetter to produce a fully uniform planet catalog, and further characterized the completeness \citep{Christiansen2020} and reliability \citep{Thompson2018} of the catalog to support statistically robust demographic studies of the Kepler planet population. Community-led Kepler automated vetting algorithms have included the deep learning classifiers \texttt{AstroNet} \citep{ShallueVanderburg2018} and \texttt{ExoMiner} \citep{Valizadegan2022}, and those designed for the Kepler follow-on mission K2 have included \texttt{AstroNet-K2} \citep{Dattilo2019}, the Discovery and Vetting of Exoplanets framework \cite[\texttt{DAVE};][]{Kostov2019}, and \texttt{EDI-Vetter} \citep{Zink2020}.

Components of TESS vetting practices involve automation, but the identification of TESS Objects of Interest (TOIs) is still heavily reliant on manual inspection. TCEs identified by the TESS Science Processing Operations Center pipeline \cite[SPOC;][]{Jenkins2016} are vetted by \texttt{TESS-ExoClass} \cite[\texttt{TEC};][]{Burke2019}, which was built off the Robovetter and \texttt{DAVE} vetting tools but is not yet fully automated. Meanwhile, signals identified by the Quick-Look Pipeline \cite[QLP;][]{Huang2020} at the TESS Science Office (TSO) have been partially triaged using the \texttt{AstroNet-Triage} machine learning classifier \citep{Yu2019} over TESS Sectors 6 -- 33 and its upgrade \texttt{AstroNet-Triage-v2} \citep{Tey2023} since Sector 34. The TESS \texttt{AstroNet} classifiers are trained to reject signals due to stellar variability and instrumental noise, but cannot yet distinguish between transiting planets and eclipsing binaries, and do not implement any pixel-level vetting for identifying off-target signals. A significant component of manual inspection of both flux- and pixel-level data by QLP operators is therefore required to further reject false positives and identify the most promising planet candidates. Signals from SPOC passing TEC and signals from QLP passing \texttt{AstroNet-Triage-v2} and operator inspection are then passed through further manual inspection to alert official TOIs \citep{Guerrero2021}.

The QLP produces light curves from Full-Frame Images (FFIs) for the $\sim20$ million stars as faint as $T = 13.5$ mag \cite[or $T = 15$ mag for M dwarfs;][]{Kunimoto2022b}, but has restricted vetting efforts to the $\sim1.5$ million stars with $T < 10.5$ mag because fainter searches would overwhelm operators with the number of candidates needing manual review. The TESS Faint Star Search \citep{Kunimoto2022} was established to identify planets orbiting the fainter stars not analyzed by QLP vetting practices by employing a Robovetter-inspired algorithm on signals passing \texttt{AstroNet-Triage}. While not fully automated, this effort has resulted in over 3400 TOIs ($\sim45\%$ of all TOIs), demonstrating the power of applying more efficient vetting algorithms to the enormous TESS dataset.

Most other TESS vetting tools are machine learning classifiers \cite[e.g.,][]{Osborn2020, Rao2021, Fiscale2021, Ofman2022, Valizadegan2022}; however, these tools are either not publicly available, do not include pixel-level vetting, are not fully automated for use on TESS data, or are only designed for subsets of the TESS dataset (e.g., 2-minute SPOC data products but not FFI data products). There remains a need for a widely applicable, fully automated TESS vetting tool, both for enabling large-scale planet searches and for producing uniform planet catalogs and occurrence rate data products for demographics.

We present \texttt{LEO-Vetter} (the Lazy-Exoplanets-Operations Vetter), a Robovetter-inspired tool for fully automated vetting of planet candidates found in TESS data. A user-friendly and pure Python version of this software is available on GitHub.\footnote{\url{https://github.com/mkunimoto/LEO-vetter} \citep{LEODOI}} We summarize the general design of \texttt{LEO-Vetter} in \S\ref{sec:leo} and describe specific flux- and pixel-level tests against noise/systematic false alarms in \S\ref{sec:noiseFPs}, and astrophysical false positives in \S\ref{sec:astroFPs}. We apply \texttt{LEO-Vetter} to $\sim200,000$ M dwarf TESS light curves and data products with simulated transits and noise properties in \S\ref{sec:validation}. In \S\ref{sec:performance} we discuss the performance of the vetter in terms of completeness, reliability, and its ability to recover known TOIs from these data products, and in \S\ref{sec:PCs} we discuss a uniformly vetted planet catalog outputted by our M dwarf search.

\section{Automated TESS Candidate Vetting with \texttt{LEO-Vetter}}\label{sec:leo}

As inputs, \texttt{LEO-Vetter} takes the time, raw flux, detrended flux, and flux uncertainties of a light curve containing a TCE to be vetted, as well as the TCE orbital period ($P$), epoch ($T$), and duration ($\tau$, as defined below). TCEs are considered signals recovered by a planet search (such as a Box-Least Squares \cite[BLS;][]{Kovacs2002} period-search algorithm) that pass specific detection criteria. There may be multiple TCEs associated with a given light curve in the case of multi-planet searches. \texttt{LEO-Vetter} will treat each of these TCEs independently, meaning that it is up to the user to resolve cases such as primary and secondary eclipses being recovered as different TCEs (e.g., see \S\ref{sec:future} for further discussion, and \S A.2 in KDR25).

The specific definition of ``duration'' may be different depending on the planet search algorithm (e.g., box- vs. trapezoid-based algorithms), but for computational efficiency the duration assumed by \texttt{LEO-Vetter} is that of a box (\S\ref{sec:SES_MES}). Therefore, when \texttt{LEO-Vetter} computes the signal-to-noise ratio (SNR) of a given TCE, durations estimated by non-BLS algorithms will result in an approximate box-like SNR.

We also assume that:

\begin{itemize}
    \item the input flux time series has already been corrected for dilution (as is the case for most publicly available TESS light curves, such as from the SPOC and QLP pipelines);
    \item the input flux time series has had outliers removed (typically performed by a user, such as by sigma-clipping);
    \item poor quality datapoints have been removed (such as by masking out observations with non-zero quality flags set by SPOC or QLP); and
    \item the transits are strictly periodic, i.e., there are no significant transit timing variations (TTVs).
\end{itemize}

\texttt{LEO-Vetter} then employs a series of both flux- and pixel-level vetting tests on the given TCE. Following the KDR25 Robovetter, the metrics produced by each test are compared to pass-fail thresholds to determine which TCEs are planet candidates (PCs), astrophysical false positives (FPs, such as eclipsing binary stars and off-target eclipsing systems), and noise/systematic false alarms (FAs, such as stellar variability, instrumental noise, and scattered light). An FA is a signal that fails any false alarm test; an FP is a signal that passes all false alarm tests, but fails at least one false positive test; and a PC is a signal that passes all tests.

The full suite of tests are described in the following sections. For each test, we provide the thresholds that resulted from our optimization described in \S\ref{sec:tuning}.  These thresholds are for a particular choice of optimization targets, and the reader may make different optimization choices resulting in different thresholds. The optimal thresholds will also depend on the specific data used in the optimization. When \texttt{LEO-Vetter} is run, users are provided with all metrics computed for a given TCE, and users can change the pass-fail thresholds applied to those metrics (or new combinations of those metrics) for their own science purposes.

\subsection{Single and Multiple Event Statistic Time Series}\label{sec:SES_MES}

Several of our vetting tests rely on estimates of the SNR of the TCE, as well as of events throughout the light curve (e.g., events shortly before or after individual transits, or potential secondary events in the phase-folded light curve). For simplicity we adopt a box-shaped transit, where the corresponding transit SNR is
\begin{equation}
    \mathrm{SNR} = \frac{\delta}{\sigma_{tr}}.\label{eqn:SNR}
\end{equation}
Here $\delta$ is the transit depth, computed as the weighted mean of all in-transit data points (entirely within half a transit duration from the transit center, however duration was defined) subtracted from the weighted mean of out-of-transit datapoints, and $\sigma_{tr}$ is the pink noise over the transit duration (see Eqn. \ref{eqn:noise}). For TESS light curves, it is crucial to weight computations involving flux by their uncertainties, given that the time series may include mixes of cadences (e.g., 30-minute FFI cadence in TESS Prime Mission vs.~200-second FFI cadence in the second Extended Mission), and different exposure times are associated with different per-point noise.

In order to account for correlated noise in the TESS light curves, we adopt the pink noise definition from \citet{Pont2006} to estimate $\sigma_{tr}$ as
\begin{equation}\label{eqn:noise}
    \sigma_{tr} = \sqrt{\frac{\sigma_{w}^{2}}{n} + \frac{\sigma_{r}^{2}}{N_{tr}}},
\end{equation}
where $\sigma_{w}$ is the white noise in the light curve, $\sigma_{r}$ is the red noise, $n$ is the number of points in-transit, and $N_{tr}$ is the number of transits. We take $\sigma_{w}$ to be equal to the weighted standard deviation of the light curve after masking out data within one transit duration of the center of the detected transits. Following \citet{HartmanBakos2016}, we estimate $\sigma_{r}$ using the expression
\begin{equation}
    \sigma_{r}^{2} = \sigma_{\mathrm{bin}}^{2} - \sigma_{\mathrm{bin,xpt}}^{2},
\end{equation}
where $\sigma_{\mathrm{bin}}$ is the weighted standard deviation of the residual light curve after binning in time with a bin-size equal to the duration of the transit, and $\sigma_{\mathrm{bin,xpt}}$ is the expected standard deviation of the binned light curve if the noise were uncorrelated in time. Following \citet{HartmanBakos2016}, in the event that $\sigma_{r}^{2}$ is estimated to be less than zero, we set $\sigma_{r} = 0$. In later analysis in this paper (\S\ref{sec:validation}), we find that $\sim30\%$ of stars in our sample end up with $\sigma_{r} = 0$, implying white noise-dominated light curves.

Aside from computing the overall SNR of the TCE to be vetted, \texttt{LEO-Vetter} measures two SNR time series, analogous to the Single Event Statistic (SES) and Multiple Event Statistic (MES) time series produced by the Kepler Transiting Planet Search module \citep{Jenkins2010}. The SES time series represents the SNR of individual signals at every cadence and is computed by considering only data within half a transit duration of a given cadence. The SES value at a given TCE transit time is equal to the SNR of that individual transit. The MES time series represents the SNR of signals at every phase after folding the light curve at the orbital period. The MES value at a phase of 0 is equivalent to the overall SNR of the TCE.
 
\subsection{Model Fits}

Several of our tests also rely on the results of model fits to the data; namely, a straight line, a trapezoid model, and a quadratic limb-darkening transit model \citep{MandelAgol2002} implemented in \texttt{batman} \citep{Kreidberg2015}. Even though a trapezoid is an approximation to a transit shape, we fit a trapezoid model because it is more robust and less prone to failing for a given TCE compared to a transit model \cite[e.g.,][]{Morton2012}. To speed up the fit process, data more than two transit durations from the TCE mid-transit times are ignored.

Because planet searches may result in tens of thousands to millions of TCEs, we use least-squares minimization with \texttt{lmfit} \citep{LMFIT} rather than a significantly more computationally intensive routine such as Markov Chain Monte Carlo \cite[MCMC;][]{Metropolis1953,Hastings1970}. Therefore, we advise users to perform their own transit model fits for planet candidates after the vetting process to produce final reported parameters.

\paragraph{Straight Line Model} The straight line model is parameterized by a simple slope and flux offset. \texttt{LEO-Vetter} records the reduced chi-square value ($\chi_{r,\mathrm{line}}^{2}$) and Akaike Information Criterion statistic (AIC$_{\mathrm{line}}$) \citep{Akaike1974} of the best fit. The purpose of fitting a straight line to a given TCE is to provide a null hypothesis against which we can test a transit model as part of automated vetting (\S\ref{sec:modelfit}).

\paragraph{Trapezoid Model} The trapezoid model is parameterized by orbital period, transit epoch, transit depth, ratio between full transit duration and orbital period ($q$), ratio between ingress duration and full transit duration ($q_{\mathrm{in}}$), and out-of-transit flux level ($f_{0}$). We fit trapezoid models in two stages: reduced-parameter fits where $q_{in}$ is set fixed to different guess values in order to improve convergence, and a final fit where all parameters are allowed to vary. Using preliminary reduced-parameter fits improves model convergence and fit consistency and is inspired by a similar strategy employed by the Kepler Data Validation module \citep{Li2019}.

The reduced-parameter fits involve setting  $q_{\mathrm{in}}$ to fixed values of 0.1, 0.15, 0.2, 0.25, 0.3, 0.35, and 0.4. All other parameters are fitted, with initial guesses determined from the TCE parameters. The best-fit parameters giving the minimum chi-square ($\chi^{2}$) are then used as the initial guess for a final fit where $q_{\mathrm{in}}$ then becomes an additional fit parameter, allowed to vary between 0 and 0.5. \texttt{LEO-Vetter} records the final best-fit parameters and their uncertainties, as well as $\chi_{r,\mathrm{trap}}^{2}$ and AIC$_{\mathrm{trap}}$.

\paragraph{Transit Model} The transit model is parameterized by orbital period, transit epoch, ratio of planet-to-star radii ($R_{p}/R_{\star}$), distance between planet and star at mid-transit in units of stellar radius ($a/R_{\star}$), impact parameter ($b$), and flux offset, with circular orbits assumed. We adopt quadratic limb-darkening parameters from \citet{Claret2017}. Similar to the trapezoid model fits, the light curve is first subjected to five reduced-parameter fits in which $b$ is set to fixed values of 0.1, 0.3, 0.5, 0.7, and 0.9. The best-fit parameters giving the minimum $\chi^{2}$ are then used as the initial guess for a final fit, where $b$ becomes an additional fit parameter, allowed to vary between 0 and 1. We cap $b$ at 1 to assist with convergence, though it means that the fits are inaccurate for highly grazing transits Users should perform their own transit model fits after running \texttt{LEO-Vetter} to improve on these parameters. \texttt{LEO-Vetter} records the final best-fit parameters and their uncertainties, as well as $\chi_{r,\mathrm{transit}}^{2}$ and $\mathrm{AIC}_{\mathrm{transit}}$.

\section{False Alarms Tests}\label{sec:noiseFPs}

Common false alarms in TESS data include intrinsic stellar noise and variability (e.g., quasi-sinusoidal light curves from pulsating stars and star spots), instrumental systematics (e.g., momentum dumps, when the spacecraft thrusters are engaged to remove angular momentum from the reaction wheels), and scattered light from the Earth and Moon (causing significant flux ramps and increased photometric scatter at the start of each TESS orbit). Tests for removing these false alarms are primarily focused on analyzing the shape and significance of events. As a reminder, the pass-fail thresholds adopted in this section were optimized using injection/recovery tests (\S\ref{sec:tuning}). However, users may set their own thresholds with \texttt{LEO-Vetter} as needed.

\subsection{SNR Test}\label{sec:recomputed}

A transit signal must be sufficiently strong against the light curve noise to be considered a promising candidate. TCEs will have already been identified based on their strength relative to some detection criteria, such as a minimum SNR value; regardless, we fail a TCE if $\mathrm{SNR} < 6.2$.

\subsection{Model Fit Test}\label{sec:modelfit}

Planet transits should be well-fit by a transit model. A TCE fails this test if \texttt{lmfit} was unable to fit a transit model to the data. A transit model should also provide a significantly better fit than a straight line. We fail a TCE if $N_{tr} \leq 10$ and $\mathrm{AIC}_{\mathrm{line}} - \mathrm{AIC}_{\mathrm{transit}} < 60$, or $N_{tr} > 10$ and $\mathrm{AIC}_{\mathrm{line}} - \mathrm{AIC}_{\mathrm{transit}} < 30$. We found that this harsher threshold for TCEs with fewer than 10 transits was needed because of the large contamination from scattered light events. Our testing indicated that the performance of this test was relatively insensitive to the exact thresholds. 

\subsection{Sine Wave Event Evaluation Test (SWEET)}\label{sec:sweet}

\begin{figure*}
\centering
    \includegraphics[width=0.48\linewidth]{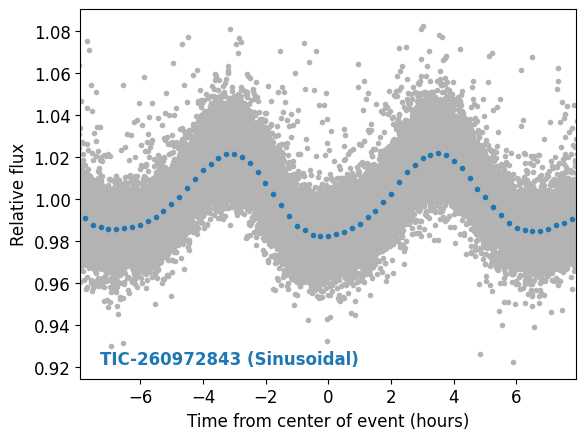}
    \includegraphics[width=0.48\linewidth]{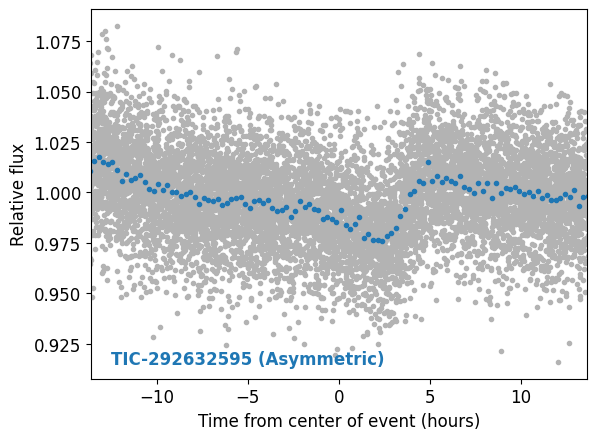}
    \includegraphics[width=0.48\linewidth]{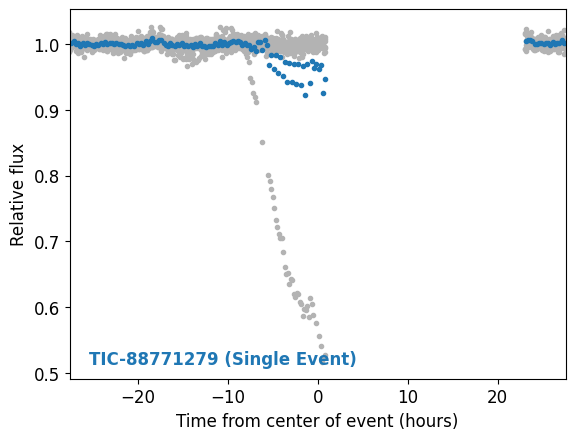}
    \includegraphics[width=0.48\linewidth]{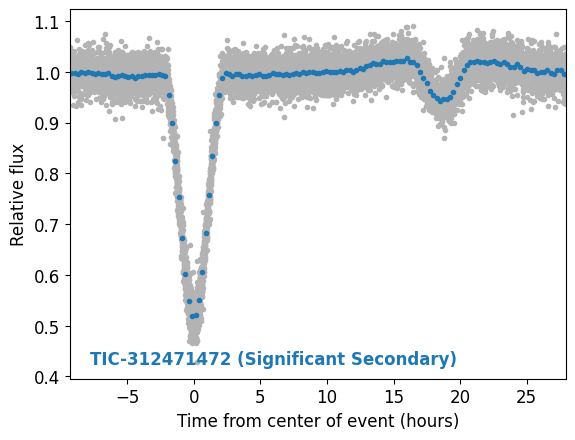}
    \caption{Examples of non-planetary TCEs to demonstrate the typical false alarms and false positives targeted by select tests in \texttt{LEO-Vetter}. We show TCEs caused by stellar variability that failed the SWEET test (\S\ref{sec:sweet}, top left) and the asymmetry test (\S\ref{sec:asym}, top right), a scattered light event that failed the single event domination test (\S\ref{sec:single}, bottom left), and an eclipsing binary that failed the significant secondary test (\S\ref{sec:secondary}, bottom right).}\label{fig:examples}
\end{figure*}

Transit searches are often confounded by stellar variability with strong quasi-periodic and sinusoidal components over short timescales, especially due to pulsating stars and star spots (e.g., top left panel of Figure \ref{fig:examples}). We adopt the Sine Wave Event Evaluation Test (SWEET) from KDR25 to identify these cases. We fit three sine waves with periods fixed to half, one, and two times the TCE period, with sine wave amplitude, phase, and flux offset allowed to vary. To avoid throwing out very deep transits, all data within one transit duration from mid-transit times are masked out. The significance of the fit is the amplitude divided by the uncertainty in the amplitude, and we record the highest significance among all three fits. We fail a TCE if it has a SWEET significance greater than 15 and $P < 10$ days.

\subsection{Unphysical Transit Duration Tests}\label{sec:unphysical}

A TCE may have a duration that is significantly longer than what is expected for a planet at a given orbital period. Most commonly, this indicates that the TCE is caused by sinusoidal oscillations or other high-frequency variability. A TCE may also have a duration that is longer than what we would expect for a planet in a circular orbit. We fail TCEs with either $q > 0.5$ or $q_{\mathrm{circ}}/q < 0.6$, where $q_{\mathrm{circ}}$ is the value of $q$ expected for a central transit in a perfectly circular orbit. We also fail TCEs with a transit model-fitted $a/R_{\star} < 1.5$, indicating that the TCE is too close to the host star to be likely planetary.

\subsection{Transit Asymmetry Test}\label{sec:asym}

Planet transits should be almost symmetric, while systematics tend to have more asymmetric or ramp-like shapes (e.g., top right panel of Figure \ref{fig:examples}). Following \citet{Eschen2024}, we fit two trapezoid models, one to only the left side of the transit and one to only the right side of the transit, with the period, epoch, and transit depth fixed to the overall best-fit trapezoid model. We quantify the asymmetry based on the significance of the difference between the left and right $q$ values,
\begin{equation}
    \mathrm{ASYM} \equiv \frac{|q_\mathrm{left} - q_\mathrm{right}|}{\sqrt{\sigma_{q_\mathrm{left}}^2 + \sigma_{q_\mathrm{right}}^2}},
\end{equation}
and fail TCEs with $\mathrm{ASYM} > 10$.

\subsection{Depth Mean-to-Median (DMM) Ratio Test}\label{sec:dmm}

The mean of all measured transit depths should be consistent with the median of all transit depths in the case of a planet transit. Thus, a depth mean-to-median (DMM) ratio significantly different from 1 can be used to identify when a candidate is due to a systematic error. We fail TCEs with DMM $>$ 1.5.

\subsection{SNR Consistency Test}\label{sec:CHI}

This test examines the SNRs of each transit individually and compares them to what is expected assuming the transit has the same depth as the overall weighted average transit depth. \texttt{LEO-Vetter} computes the expected SNR of the $i$th transit ($\langle \mathrm{SNR}_{i} \rangle$) from Eqn. \ref{eqn:SNR}, with the transit depth set to the overall transit depth, and the actual SNR of the $i$th transit ($\mathrm{SNR}_{i}$) with transit depth set to the actual $i$th transit depth. We compare the two SNR estimates using the $\chi^{2}$ statistic,
\begin{equation}
    \chi_{\mathrm{SNR}}^{2} \equiv \sum_{i=1}^{N_{tr}}\frac{(\mathrm{SNR}_{i} - \langle\mathrm{SNR}_{i}\rangle)^{2}}{\langle\mathrm{SNR}_{i}\rangle},
\end{equation}
and define the CHI metric as
\begin{equation}
    \mathrm{CHI} \equiv \mathrm{SNR} \times \bigg(\frac{\chi_{\mathrm{SNR}}^{2}}{N_{tr} - 1}\bigg)^{-1/2}.
\end{equation}
We fail TCEs with $\mathrm{CHI} < 7.8$.

\subsection{Chases Test}\label{sec:chases}

As described in KDR25, the Kepler team developed the ``Chases'' metric to identify non-transit-like, low SNR events that do not stand out against the local noise in the light curve. This metric makes use of the full SES time series. We first find the SES for an individual transit ($\mathrm{SES}_{i}$) as the largest SES over the in-transit cadences. The SES time series is then searched for $\Delta t$, the time of the closest signal with $|\mathrm{SES}| > 0.7~\mathrm{SES}_{i}$. The search range starts at 1 transit duration from midtransit, up to a maximum of $\Delta t_{\mathrm{max}} = P/10$, on either side of the event. The final Chase metric is determined as ${C}_{i}=\min (\Delta t, \Delta t_{\mathrm{tmax}})/\Delta t_{\mathrm{max}}$. A value if $C_{i} \approx 0$ indicates an event of comparable strength is close to the transit event, while a value of $C_{i} = 1$ indicates that the transit is unique. Chases metrics are only computed for TCEs with five or fewer transit events.

We fail TCEs with mean of individual Chases metrics of less than 0.78.

\subsection{Uniqueness Tests}\label{sec:uniqueness}

Planet transits should be unique compared to other events in the folded light curve with the same duration. To quantify uniqueness we adopt the Model-Shift Uniqueness Test developed for Kepler \citep{Coughlin2017}, which has been used in several automated vetting pipelines and codebases such as \texttt{DAVE}, \texttt{TEC}, and the Faint Star Search.

We refer interested readers to \citet{Coughlin2017} for the full details of the Model-Shift test. In summary, Model-Shift uses a transit model as a template to measure the amplitude of transit-like events at all phases in the phase-folded light curve. The procedure measures the significance and phases of the primary transit event, secondary and tertiary events, and the most significant positive flux (inverted transit-like) event, where significance is defined as the amplitude divided by the uncertainty in the amplitude. Model-Shift also estimates the threshold at which a signal is statistically significant under the assumption of Gaussian noise ($\mathrm{FA}_{1}$), the threshold at which the difference in significance between two events can be considered statistically significant ($\mathrm{FA}_{2}$), and the ratio of systematic red noise at the duration of the transit to the white noise in the light curve ($F_{\mathrm{red}}$).

For our Python implementation of Model-Shift, we reduce computational expense by using the box-model transit depths and MES time series to quantify the amplitude and significance of transit-like events, respectively, rather than the convolved transit model. This is similar to the uniqueness test procedure outlined by \citet{Zink2020}. We measure the significance of the primary transit event as the largest MES value in the MES time series ($\mathrm{MES}_{\mathrm{pri}}$). The most significant event at least two transit durations from the primary is labeled the secondary event ($\mathrm{MES}_{\mathrm{sec}}$), and the next most significant event at least two transit durations from both the primary and secondary is labeled the tertiary event ($\mathrm{MES}_{\mathrm{ter}}$). Finally, the most significant positive event ($\mathrm{MES}_{\mathrm{pos}}$) at least three transit durations from the primary and secondary is also labeled.

We use the following quantities as decision metrics:

\begin{equation}
\begin{split}
    \mathrm{MS}_{1} &\equiv \mathrm{MES}_{\mathrm{pri}}/F_{\mathrm{red}} - \mathrm{FA}_{1}\\
    \mathrm{MS}_{2} &\equiv (\mathrm{MES}_{\mathrm{pri}} - \mathrm{MES}_{\mathrm{ter}}) - \mathrm{FA}_{2} \\
    \mathrm{MS}_{3} &\equiv (\mathrm{MES}_{\mathrm{pri}} - \mathrm{MES}_{\mathrm{pos}}) - \mathrm{FA}_{2}.
\end{split}
\end{equation}

\noindent We fail TCEs with $\mathrm{MS}_{1} < 0.2$ (i.e., the TCE is too similar to the systematic noise of the light curve), $\mathrm{MS}_{2} < 0.8$ (the TCE is too similar to the tertiary event), or $\mathrm{MS}_{3} < 0.8$ (the TCE is too similar to the most positive event).

This test may fail TCEs associated with planets in resonant systems, where additional planets may leave  features with high MES in the phase-folded time series.

\subsection{Transit Shape Test}\label{sec:shape}

An additional output of Model-Shift is the $\mathrm{SHP}$ metric, which determines if the measured amplitudes deviate from the mean value more in the positive flux direction, negative flux direction, or are symmetrically distributed in both directions. The $\mathrm{SHP}$ metric is computed from 
\begin{equation}
    \mathrm{SHP} \equiv \frac{F_{\mathrm{max}}}{F_{\mathrm{max}} - F_{\mathrm{min}}},
\end{equation}
where $F_{\mathrm{max}}$ and $F_{\mathrm{min}}$ are the maximum and minimum measured flux amplitudes, respectively. $\mathrm{SHP}$ lies between 0 and 1, where 0 indicates the light curve only decreases in flux, consistent with a planet transit, and a value near 1 indicates the light curve only increases in flux, such as for a lensing event or systematic outlier. We fail TCEs with $\mathrm{SHP} > 0.6$.

\subsection{Single Event Domination Test}\label{sec:single}
TCEs caused by scattered light or detrending issues near the edges of TESS data gaps are often caused by a single deep outlier event (e.g., bottom left panel of Figure \ref{fig:examples}). If the largest SNR value of a TCE's transit events (SNR$_{\mathrm{max}}$) divided by the overall TCE SNR value is large, this indicates that one of the individual events dominates when calculating SNR. We fail TCEs with ten or fewer transits and $\mathrm{SNR}_{\mathrm{max}} > 0.88~\mathrm{SNR}$.

\subsection{Data Gap Test}\label{sec:gaps}

While the Single Event Domination Test is able to remove many scattered light events corresponding to single outliers, scattered light events may also manifest as more consistent transits near the 13.7-day TESS orbit. \texttt{LEO-Vetter} counts the number of transits near gaps (i.e., has a transit mid-center time within 2 transit durations of a gap), where a gap is defined as at least 0.3 days of missing data, corresponding roughly to the typical time between orbits. We fail TCEs if 50\% or more of the associated transits are near gaps.

\subsection{Individual Transit Tests}\label{sec:indiv}

The previous tests primarily assess the shape and significance of the TCE in the phase-folded light curve. However, events can individually be non-transit-like and only appear transit-like when folded together. The following tests flag individual transits as non-transit-like. We require that a TCE still has $N_{tr} \geq 3$ and $\mathrm{SNR} > 6.2$ after throwing out all bad transits.

The majority of these individual transit tests were adopted from KDR25, as described below. The only tests we do not implement were those that targeted Kepler-specific false alarms: the ``Skye'' test, which handled  rolling-band image artifacts, and the ``Tracker'' test, which was closely tied to the interaction between the TPS and DV modules of the Kepler pipeline.

\paragraph{SES Artifacts} For TCEs with five or fewer transits, we flag individual events with a Chases metric (\S\ref{sec:chases}) $C_{i} < 0.01$.

\paragraph{Negative Significance} Transits should only correspond to decreases in flux. Following the ``Zuma'' test described by KDR25, we flag any transit with $\mathrm{SNR}_{i} < 0$, corresponding to a flux increase.

\paragraph{Poor Transit Shape} Individual transits should be well fit by a consistent transit model. We compute the reduced chi-square value for each individual transit ($\chi_{r,i}^{2}$) by comparing the flux within two transit durations of the transit center to the overall best-fit transit model, and flag any transit if it has $\chi_{r,i}^{2} > 5$. This is similar in motivation to the Kepler ``Marshall'' test; however, Marshall involved fitting individual  events to transit models and performing model comparison with alternative noise-like models, whereas we simply assess each individual transit's goodness of fit considering the overall best-fit transit model. We found that this was more effective for handling transits with very low SNR, where new model fits often failed.

\paragraph{Missing Data} TCEs commonly include events which are missing a significant amount of data either before, during, or after the transit. These often correspond to events at the edge of data gaps where strong scattered light events or detrending artifacts can trigger BLS. Similar to the ``Rubble'' test described by KDR25, for each individual transit, we count the number of data points within two transit durations of the transit center and compare it to the expected number of data points given the cadence of the observations. Importantly, we take into account the cadence of the local light curve (estimated as the median difference in time between datapoints within two durations of the transit center), given that the input light curves may include different cadences from different stages of the TESS mission. A transit is flagged if it is missing more than 25\% of the expected cadences.

\section{Astrophysical False Positive Tests}\label{sec:astroFPs}

Astrophysical false positives include eclipsing binary stars (EBs), either grazing the target star or being part of a nearby eclipsing binary (NEB) system that is contaminating the TESS photometry. We also classify planets that transit stars other than the target as astrophysical false positives. Many EBs can be quickly identified by eclipse depths too deep to be consistent with a planet transit, secondary events in the light curve, or significant odd and even event depth differences which suggest an EB was found with half the correct orbital period. Meanwhile, NEBs and off-target planets are further identified through pixel analysis.

As with the tests against false alarms, these tests are computed on all TCEs.

\subsection{Candidate Size Test}\label{sec:size}

Perhaps the most telltale sign of an EB is a transit depth that is too deep to be planetary. We multiply the stellar radius by the transit model-fitted $R_{p}/R_{\star}$ to estimate planet radius, and fail TCEs with $R_{p} > 22~R_{\oplus}$. This is a generously high radius threshold especially for long-period planets which are unlikely to be inflated; however, the planet parameters estimated by our least-squares transit model fits (especially for grazing planets) can be uncertain. More comprehensive transit modeling post-\texttt{LEO-Vetter} can further reduce contamination of stellar objects.

\subsection{V-Shaped Test}\label{sec:vshape}

While some EBs can be readily identified by their large sizes, sufficiently grazing EBs can have shallow eclipse depths that appear transit-like. Here, we adopt the V-Shape metric from KDR25, which required passing TCEs to have $V = R_{p}/R_{\star} + b < 1.04$. However, we use a more lenient threshold of $V < 1.5$ to avoid failing high-impact-parameter gas giants, especially those transiting small stars.

\subsection{Significant Secondary Test}\label{sec:secondary}

A secondary eclipse could manifest as a significant secondary event in the phased light curve (e.g., bottom right panel of Figure \ref{fig:examples}). We use the following quantities as decision metrics to determine the significance of the secondary, following the results of the uniqueness test:
\begin{equation}
\begin{split}
    \mathrm{MS}_{4} &\equiv \mathrm{MES}_{\mathrm{sec}}/F_{\mathrm{red}} - \mathrm{FA}_{1}\\
    \mathrm{MS}_{5} &\equiv (\mathrm{MES}_{\mathrm{sec}} - \mathrm{MES}_{\mathrm{ter}}) - \mathrm{FA}_{2} \\
    \mathrm{MS}_{6} &\equiv (\mathrm{MES}_{\mathrm{sec}} - \mathrm{MES}_{\mathrm{pos}}) - \mathrm{FA}_{2}.
\end{split}
\end{equation}
A secondary is considered significant if $\mathrm{MS}_{4} > 0$, $\mathrm{MS}_{5} > -1$, and $\mathrm{MS}_{6} > -1$.

Occasionally, we found that TESS light curves with very deep secondary eclipses could have unusually high $F_{\mathrm{red}}$ values because the eclipses are clear variations from Gaussian noise. As a result, the first condition ($\mathrm{MS}_{4} > 0$) would not be met and the secondary would be erroneously considered insignificant. We therefore ignore this first condition if $F_{\mathrm{red}} > 1.8$, and rely on the other conditions to flag significant secondary events.

Meanwhile, significant secondary events may not necessarily be confirmation of an EB. Some giant planets close to their stars, such as hot Jupiters, can have eclipses due to planetary occultations via reflected light and thermal emission. A TCE is allowed to pass this test if the secondary eclipse depth is less than 10\% of the primary transit depth, the impact parameter of the primary transit is less than 0.95, the derived planet radius is less than $22~R_{\oplus}$, and the planetary albedo required to produce the secondary is less than 1. We estimate the geometric albedo using 
\begin{equation}
    A = \delta_{\mathrm{sec}}\bigg(\frac{a}{R_{p}}\bigg)^{2},
\end{equation}
where $\delta_{\mathrm{sec}}$ is the depth of the secondary event.

\subsection{Odd-Even Test}\label{sec:oddeven}

Secondary eclipses can be marked as half of the primary events if an EB is detected at half its actual orbital period, and its eclipses would otherwise occur near phase 0.5. These EBs can be identified as candidates with significantly different odd and even transit depths.

We quantify the significance of the odd-even difference in several ways. First, we find
\begin{equation}\label{eqn:OE}
    \mathrm{OE}_{\mathrm{box}} = \frac{|\delta_\mathrm{odd} - \delta_\mathrm{even}|}{\sqrt{\sigma_{\delta_\mathrm{odd}}^2 + \sigma_{\delta_\mathrm{even}}^2}}, 
\end{equation}
where $\delta_\mathrm{odd}$ and $\delta_\mathrm{even}$ are the mean depths of odd and even transits, respectively, and $\sigma_\mathrm{odd}$ and $\sigma_\mathrm{even}$ are the noise over the corresponding transits. We fail TCEs with $\mathrm{OE}_{\mathrm{box}} > 3$.

Because this test implicitly assumes a box shape, we also fit trapezoid models to odd and even transits separately, where only $\delta$ and $T$ are allowed to vary and all other parameters are fixed to the overall best-fit trapezoid model values. We quantify the significance of the difference in transit depths as 
\begin{equation}
    \mathrm{OE}_{\mathrm{trap}} \equiv \frac{|\delta_\mathrm{trap,odd} - \delta_\mathrm{trap,even}|}{\sqrt{\sigma_{\delta_\mathrm{trap,odd}}^2 + \sigma_{\delta_\mathrm{trap,even}}^2}},
\end{equation}
where the $\sigma$ values correspond to the $\delta$ fit uncertainties. We also quantify the significance of the difference between transit times as 
\begin{equation}
    \mathrm{OE}_{\mathrm{trap},T} \equiv \frac{|T_\mathrm{odd} - T_\mathrm{even}|}{\sqrt{\sigma_{T_\mathrm{odd}}^2 + \sigma_{T_\mathrm{even}}^2}}.
\end{equation} 
This metric is useful for flagging cases where the primary and secondary eclipses are not exactly separated by half an orbit (i.e., the EB is in a slightly eccentric orbit), but are still close enough to be detected as odd and even transits.

Finally, we repeat this process for the transit model where only $R_{p}/R_{\star}$ and $T$ are allowed to vary. Analogously, this provides $\mathrm{OE}_{\mathrm{transit}}$ and $\mathrm{OE}_{\mathrm{transit}, T}$. We fail TCEs with either $\mathrm{OE}_{\mathrm{trap}} > 3$ or $\mathrm{OE}_{\mathrm{transit}} > 3$, indicating significant depth differences. We also fail TCEs with $\mathrm{OE}_{\mathrm{trap},T} > 10$ or $\mathrm{OE}_{\mathrm{transit},T} > 10$, indicating significant transit time differences.

All three approaches to identifying significant odd-even depth differences (via box, trapezoid, and transit models), and the two approaches to identifying odd-even timing differences (via trapezoid and transit models) are designed to remove the same type of astrophysical false positive: EBs in near-circular orbits identified at half the correct the period. However, we found that the trapezoid and transit model fits with \texttt{lmfit} occasionally failed to produce uncertainties, leaving the vetter unable to estimate the significance of the difference in depth or times. Using all three sets of tests in combination was the most conservative approach to cover cases where one or more tests failed.

\subsection{Centroid Offset Test}\label{sec:centroid}

A powerful method for the identification of NEBs and nearby transiting systems is the difference image technique described by \citet{Bryson2013}. \texttt{LEO-Vetter} employs the \texttt{transit-diffImage} package\footnote{\url{https://github.com/stevepur/transit-diffImage}} to produce difference images and perform a Pixel Response Function (PRF) fit to measure the centroid offset. In summary, the average of in- and out-of-transit pixels surrounding the target are found from a TCE's ephemerides and duration, for each individual transit. The out-of-transit image represents a direct image of the field surrounding the target star, when no flux changes are occurring on average. These direct and difference images are produced for each TESS sector by averaging across all individual transits observed in the given sector.  The difference between in- and out-of-transit images should appear starlike at the location of the transit source, assuming the transit is the explanation for any difference in flux. We then determine the location of the transit signal source from the difference image by using the TESS PRF to a create synthetic image of a star at the location that best replicates the difference image.  We ``fit'' the transit location by varying the position of the PRF used to create the synthetic image until it minimized the RSS difference between the synthetic image and the difference image. This fit is performed by the \texttt{tess\_PRF\_centroid} method of the \texttt{transitCentroids} class in \texttt{transit-diffImage}.  When the transit SNR is high, this fitting method is accurate to within about 0.2 pixels ($= 5.2$ arcsec with TESS' 21-arcsec pixels), compared with Kepler ($= 0.8$ arcsec with Kepler's smaller 4-arcsec pixels) \citep{Bryson2025}, but this accuracy degrades with lower transit SNR as the difference image looks less and less like a single star.  The \texttt{tess\_PRF\_centroid} method returns a quality metric \texttt{prfFitQuality} measuring how well the difference image matches a single star, with \texttt{prfFitQuality} = 1 being a perfect match.  We recommend weighting any analysis by this quality metric, and \texttt{prfFitQuality} $< 0.3$ means that the fit results should not be trusted.

We compute the pixel position of stars from their catalog RA and Dec using the python \texttt{tess-point} package \citep{tesspoint}, which accounts for aberration when converting from (RA, Dec) to pixel position.  However the inverse conversion from pixel position to (RA, Dec) in \texttt{tess-point} does not account for aberration, so we use the \texttt{pix\_to\_ra\_dec} function in \texttt{transit-diffImage}, which finds the (RA, Dec) that \texttt{tess-point} converts to the desired pixel position via nonlinear fitting.  In this way, we find the (RA, Dec) of the location of the transit signal from the difference image, and we define $\Delta \theta \equiv $ the distance of the transit signal's (RA, Dec) from the target star's catalog (RA, Dec).   

\begin{figure*}
    \includegraphics[width=\linewidth]{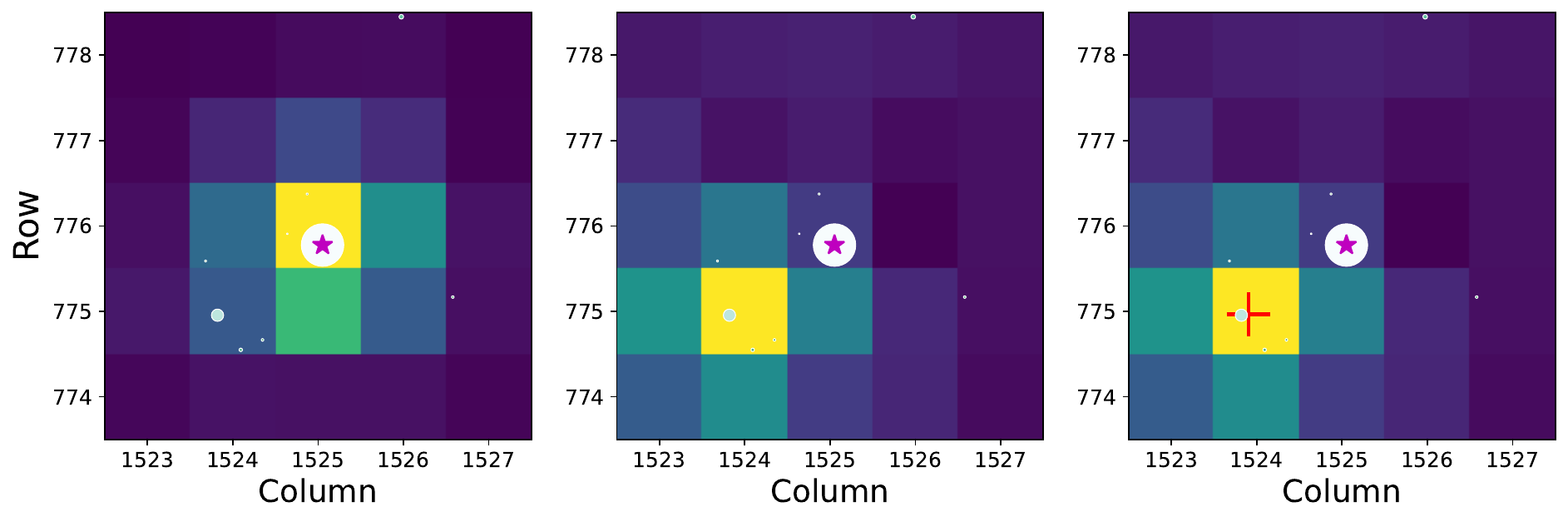}
    \caption{An example of difference image analysis of the known false positive TOI-164.01. Left: average out-of-transit pixel image showing the pixels near TOI-164.01 (the average in-transit image looks very similar).  Stars are shown as white-bordered disks with size and color determined by their TIC magnitudes, and the target star TOI-164 is marked by the magenta '$\star$'. Center: the difference pixel image, showing that the largest change from out-of-transit to in-transit is on the pixels containing the brightest background star.  Visually, the star-like pattern in the difference image looks like it is due to that brightest background star.  Right: the location of the PRF fit to the difference image is shown as the red '$+$', demonstrating that the PRF fit indicates that the transit signal is on the brightest background star. The colors scales are different between the images, but the quantitative details are irrelevant.}\label{fig:diff_im_offset_example}
\end{figure*}

We fail TCEs as likely off-target when $\Delta\theta > 15^{\prime\prime}$. For stars with multiple sectors of data, we set $\Delta\theta$ to the offset measured with the highest fit quality. We also experimented with defining $\Delta\theta$ as the mean of all offsets across sectors weighted by fit quality, but found that this occasionally resulted in on-target signals being incorrectly flagged as off-target due to poor-quality fits. We favored high completeness over high reliability for this triage step, given that more robust but computationally expensive analyses (e.g., positional probability) can still be used post-\texttt{LEO-Vetter} to reduce false positive contamination.

Because of the time and computational expense of downloading FFI cutouts and producing difference images, we recommend only performing this test for TCEs passing all flux-level tests.

\section{Testing on Observed and Simulated Data}\label{sec:validation}

Ideally, our vetting tool will successfully classify all planets as PCs and all false alarms and false positives as non-PCs. Realistically, sacrifices must be made to maintain a balance between recovering many real planets while keeping the contamination from false alarms and false positives low. We searched and applied our vetter on three different datasets to assess different aspects of its performance:

\begin{enumerate}
    \item Multi-sector light curves obtained by TESS, in order to explore the characteristics of real signals passing the vetter and to see how they compare to known TOIs. We also used this dataset to hand-tune pass/fail thresholds against astrophysical false positives.
    \item Transits injected into the TESS light curves, in order to simulate planets in the data and test the vetter's ability to classify planets as PCs.
    \item TESS light curves with re-orderings of both orbits and sectors, in order to simulate false alarms in the data and test the vetter's ability to classify such signals as FAs.
\end{enumerate}

\subsection{Observed TESS Light Curves}

For our test observations, we selected 198,640 M dwarfs ($R_{*} < 0.6~R_{\odot}$, $M_{*} < 0.6~M_{\odot}$, $2400 < T_{\mathrm{eff}} < 3900 K$) brighter than $T = 14$ mag and with an estimated contamination ratio $< 0.5$ from the TESS Input Catalog (TIC) Candidate Target List (CTL) v8.01 \citep{Stassun2019}. We chose M dwarfs as these have typically noisy light curves due to stellar activity and the relative faintness of their hosts. Of these M dwarfs, 172,486 have FFI light curves produced by the QLP in Sectors 1 -- 70. The QLP dataset provides the longest coverage across multiple TESS sectors compared to other publicly available datasets, allowing us to test \texttt{LEO-Vetter} on light curves with unique mixes of cadences (30-minute, 10-minute, and 200-second), observing baselines, and numbers of sectors and data gaps.

We removed QLP observations with poor quality (non-zero) quality flags and normalized each light curve to 1 before combining individual sectors into a multi-sector light curve. Each light curve was then detrended using the biweight time-windowed slider implemented in the \texttt{wotan} package \citep{Hippke2019} with a window of 0.5 days to remove astrophysical variability longer than the timescales of typical M dwarf planet transits (a few hours).

\subsection{Observed TCEs}

We searched for signals with periods between 1 day and a target-dependent maximum orbital period ($P_{\mathrm{max}}$) using the BLS routine implemented in the \texttt{cuvarbase} time series analysis package \citep{Hoffman2022}. To account for the fact that TESS data often features multiple gaps and that different stars can have different numbers of sectors, we set $P_{\mathrm{max}}$ to half the length of the longest continuous stretch of consecutive sectors for a given star, or at most 40 days in order to avoid increases in false alarms due to scattered light events. The median $P_{\mathrm{max}}$ was 13.5 days while the median total observing baseline was 758 days, indicating that most stars in our sample were observed for only one sector at a time with two-year-long data gaps. 

For simplicity, we only searched each light curve once, returning one TCE per star. Defining a TCE as a BLS-detected signal with SNR $> 9$ and $N_{tr} \geq 3$, we uncovered 18,424 observed TCEs.

The original detrending process can be destructive to the shape of a planetary transit and often results in a loss in SNR. We re-detrended each TCE's light curve by setting the detrending algorithm to ignore cadences within 1 duration of the TCE mid-points. These re-detrended light curves were later provided as inputs to \texttt{LEO-Vetter} for TCE vetting.

\subsection{Simulated Planet TCEs}

We injected one planet into every pre-detrended light curve with orbital periods uniformly drawn over $P \in (1, P_{\mathrm{max}})$ days, planet radii log-uniformly drawn over $R_{p} \in (0.5, 16)~R_{\oplus}$, impact parameters uniformly drawn over $b \in (0, 1)$, and transit epochs uniformly drawn in orbital phase. Each transit was created using a quadratic limb-darkening model \citep{MandelAgol2002} assuming circular orbits. A planet was redrawn if it did not result in at least 3 transits in the data; otherwise, it would not meet our detection criteria and therefore not make its way to vetting.

We then detrended, searched, and re-detrended the light curves using the same procedure as the original data. To match BLS detections with injected signals, we checked if the BLS periods and epochs matched the injected periods and epochs with significance $\sigma_{P} > 2.5$ and $\sigma_{T} > 2$, as defined by \citet{Coughlin2014}. We recovered 62,799 injected TCEs.

\subsection{Simulated False Alarm TCEs}

The simulated false alarm data should result in the detection of realistic signals with noise properties similar to the observed data while avoiding the detection of true exoplanets. The KDR25 strategy was to produce a scrambled dataset where Kepler quarters were re-ordered, thus breaking the ephemerides of real transits, as well as an inverted dataset where positive and negative flux values were switched, thus ensuring that any inverted TCE corresponded to a non-planetary, positive flux event in the original light curves. The combination of both scrambled and inverted datasets was able to largely reproduce the period distribution of false alarms seen in Kepler data.

\begin{figure}
    \includegraphics[width=\linewidth]{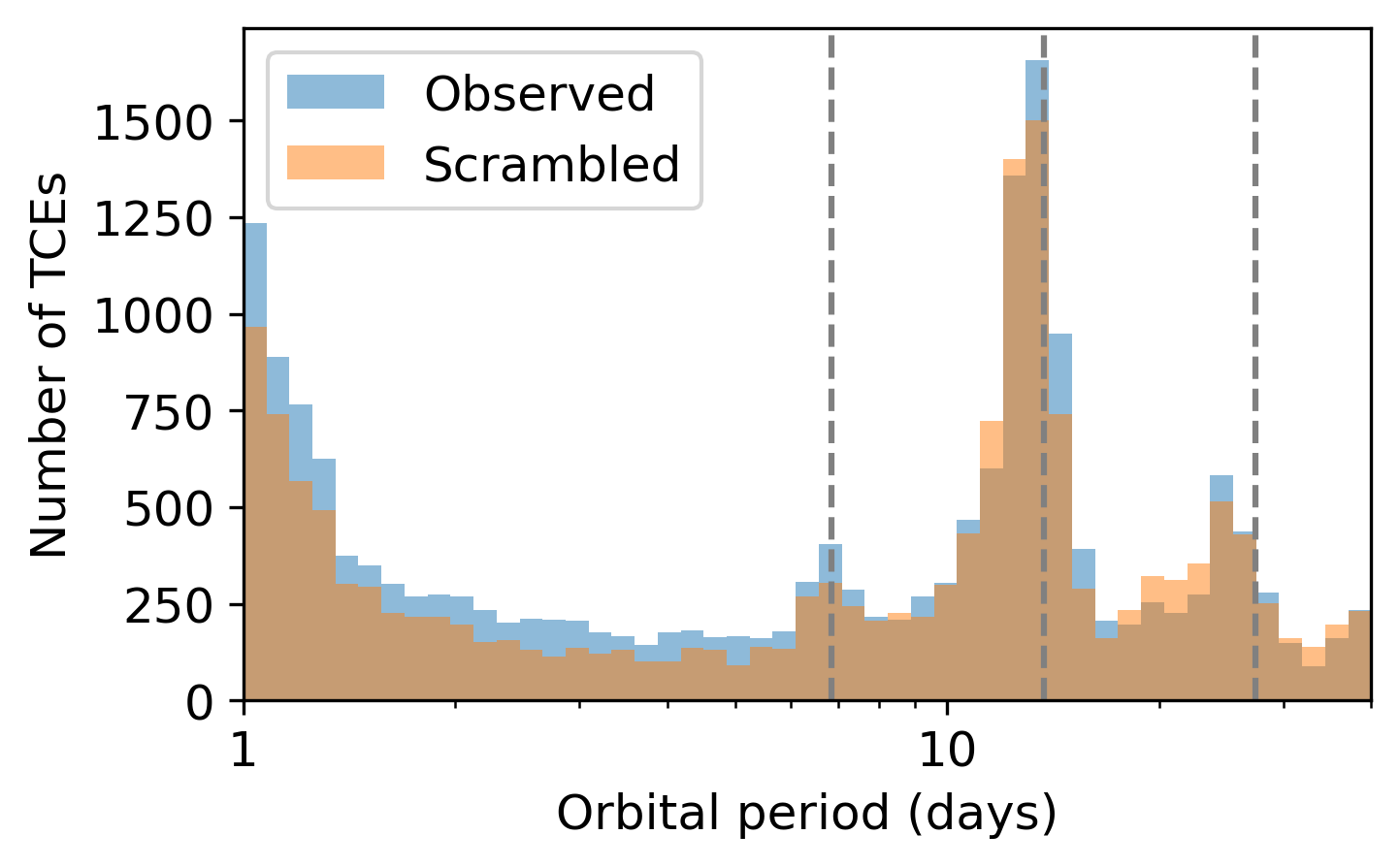}
    \caption{Period histograms of TCEs from observed (un-scrambled) light curves (blue) and scrambled light curves (orange), demonstrating that scrambling effectively simulates false alarm TCEs. The dotted grey lines indicate half, one, and two times the TESS orbital period (13.7 days). These histograms also demonstrate that a large majority of TESS TCEs are false alarms. However, \texttt{LEO-Vetter} is capable of producing planet catalogs with very high false alarm reliability (\S\ref{sec:flux}).}\label{fig:tce_hist}
\end{figure}

For TESS, we chose to produce only scrambled versions of our light curves. First, we swapped the order of TESS orbits within each sector to make sure that stars with single sectors could be scrambled, while still maintaining false alarm characteristics on orbit timescales. Then, we randomly swapped the order of all TESS sectors in the data. Each light curve was detrended, searched, and re-detrended using the same process as for the observed and injected sets, resulting in 16,134 scrambled TCEs. As seen in Fig. \ref{fig:tce_hist}, scrambling is able to reproduce the key aspects of the period distribution of observed (un-scrambled) TCEs.

We initially considered light curve inversion, but found that the most problematic TESS false alarms were due to scattered light which is not symmetric under inversion. Our inverted TCEs were also contaminated by a significant number of flare-like events, especially prevalent in M dwarf light curves, which did not reflect realistic false alarms in the original data.

\subsection{Cleaning Simulated TCEs}

Deep eclipses and transits present in the original data may affect the vetter's performance on simulated data. For example, such signals may be picked up in the scrambled dataset as planet-like signals (albeit with broken ephemerides) while not representing the types of false alarms we aimed to simulate. We removed 389 and 479 signals from the injected and scrambled TCE list, respectively, which corresponded to light curves with either known TOIs, potential eclipsing binaries listed in the TESS EB Catalog \citep{Prsa2022}, or signals that we determined to be previously unknown planet candidates or eclipsing binaries manually identified while testing the vetter. This left 62,410 injected and 15,655 scrambled TCEs. Overall, the cleaning process removed only 0.62\% injected and 2.97\% scrambled TCEs.

\subsection{Tuning Pass-Fail Thresholds}\label{sec:tuning}

The observed, injected, and scrambled TCEs were run through \texttt{LEO-Vetter} to produce a set of metrics for each TCE. Because we simulated planets and false alarms at the flux level and not the pixel level, we only applied flux-level tests. We aimed to find pass-fail thresholds that maximized both the completeness and false alarm reliability of the observed planet catalog. Here, completeness refers to the fraction of detected planets that are successfully classified as PCs, while reliability refers to the fraction of PCs that are not false alarms.

Following KDR25, completeness ($C$) is estimated by dividing the number of injected TCEs classified as PCs ($N_{\mathrm{PC}_{\mathrm{inj}}}$) by the total number of injected TCEs ($N_{\mathrm{inj}}$),
\begin{equation}
    C \approx \frac{N_{\mathrm{PC}_{\mathrm{inj}}}}{N_{\mathrm{inj}}}.
\end{equation}
Following \citet{Bryson2020prob}, effectiveness ($E$) is defined as the fraction of false alarms correctly classified as FAs, and is estimated by dividing the number of scrambled TCEs classified as FAs ($N_{\mathrm{FA}_{\mathrm{scr}}}$) by the total number of scrambled TCEs ($N_{\mathrm{scr}}$),
\begin{equation}
    E \approx \frac{N_{\mathrm{FA}_{\mathrm{scr}}}}{N_{\mathrm{scr}}}.
\end{equation}
Finally, reliability against false alarms ($R_{\mathrm{FA}}$) is estimated as \cite[Eqn. 7 from][]{Bryson2020prob},
\begin{equation}
    R_{\mathrm{FA}} \approx 1 - \frac{N_{\mathrm{FA}_{\mathrm{obs}}}}{N_{\mathrm{PC}_{\mathrm{obs}}} + N_{\mathrm{FP}_{\mathrm{obs}}}}\bigg(\frac{1 - E}{E}\bigg),\label{eqn:reliability}
\end{equation}
where $N_{\mathrm{FA}_{\mathrm{obs}}}$, $N_{\mathrm{FP}_{\mathrm{obs}}}$, and $N_{\mathrm{PC}_{\mathrm{obs}}}$ are the numbers of FAs, FPs, and PCs identified among observed TCEs, respectively.

We ran an optimization routine by varying pass-fail thresholds and computing the resulting completeness and reliability values, with the goal of minimizing the function $\sqrt{(1 - C)^{2} + (1 - R_{\mathrm{FA}})^{2}}$. Starting with a guess set of thresholds, the function was minimized using the \texttt{differential\_evolution} algorithm provided in \texttt{scipy.optimize}. Because this exercise is designed to characterize reliability against false alarms, we let the optimizer vary only a subset of thresholds corresponding to tests against false alarms (\S\ref{sec:noiseFPs}) as indicated in Table \ref{tab:results}. Tests designed to remove astrophysical false positives (\S\ref{sec:astroFPs}) had fixed thresholds based on testing the vetter on manually identified EBs.

\begin{table*}[t!]
    \centering
    \begin{tabular}{l|c|ccc}
    \hline\hline
        Flux-Level False Alarm Tests & Optimized? & Observed TCEs & Injected TCEs & Scrambled TCEs \\
        & & Failed (\%) & Failed (\%) & Failed (\%) \\
    \hline
        SNR Test (\S\ref{sec:recomputed}) & Y & 23.63 & 0.08 & 27.35 \\
        Model Fit Test (\S\ref{sec:modelfit}) & Y & 30.91 & 1.64 & 34.09 \\
        Sine Wave Event Evaluation Test (\S\ref{sec:sweet}) & Y & 4.58 & 0.01 & 3.15 \\
        Unphysical Transit Duration Tests (\S\ref{sec:unphysical}) & N & 64.32 & 0.22 & 63.88 \\
        Transit Asymmetry Test (\S\ref{sec:asym}) & Y & 0.91 & 0.00 & 0.36 \\
        Depth Mean-to-Median (DMM) Ratio Test (\S\ref{sec:dmm}) & Y & 25.76 & 0.08 & 29.41 \\
        SNR Consistency Test (\S\ref{sec:CHI}) & Y & 63.06 & 0.58 & 73.76 \\
        Chases Test (\S\ref{sec:chases}) & Y & 31.59 & 1.42 & 37.82 \\
        Uniqueness Tests (\S\ref{sec:uniqueness}) & Y & 25.18 & 0.29 & 26.48 \\
        Transit Shape Test (\S\ref{sec:shape}) & Y & 1.87 & 0.02 & 1.88 \\
        Single Event Domination Test (\S\ref{sec:single}) & Y & 39.82 & 0.29 & 45.74 \\
        Data Gap Test (\S\ref{sec:gaps}) & Y & 16.33 & 0.26 & 16.77 \\
        Individual Transit Tests (\S\ref{sec:indiv}) & N & 50.97 & 3.27 & 57.39 \\
        \hline
        Total (failed at least one false alarm test) & - & 97.13 & 6.29 & 99.91 \\
    \hline\hline
        Flux-Level Astrophysical False Positive Tests & Optimized? & Observed TCEs & Injected TCEs & Scrambled TCEs \\
        & & Failed (\%) & Failed (\%) & Failed (\%) \\
    \hline
        Candidate Size Test (\S\ref{sec:size}) & N & 23.05 & 1.28 & 24.45 \\
        V-Shaped Test (\S\ref{sec:vshape}) & N & 0.51 & 0.23 & 0.64 \\
        Significant Secondary Test (\S\ref{sec:secondary}) & N & 4.01 & 0.12 & 3.23 \\
        Odd-Even Tests (\S\ref{sec:oddeven}) & N & 17.38 & 1.65 & 17.29 \\
        \hline
        Total (failed at least one false positive test) & - & 38.99 & 3.19 & 39.55 \\
        \hline\hline
        All Flux-Level Tests & - & 98.24 & 9.00 & 99.92 \\

    \end{tabular}
    \caption{The failure rate of all flux-level tests incorporated into our fully automated vetting tool called \texttt{LEO-Vetter}, designed to reject both false alarms and astrophysical false positives. Planet candidates are signals that pass all tests. The ``Optimized?'' column indicates which tests had pass-fail thresholds fine-tuned using an automated optimization scheme. Tests not automatically optimized had thresholds manually fixed to values based on testing the vetter. The thresholds giving these results were chosen to maximize both completeness and reliability against false alarms, but other thresholds can provide higher completeness or higher reliability as needed. \texttt{LEO-Vetter} also includes pixel-level vetting through a centroid offset test (\S\ref{sec:centroid}).}
    \label{tab:results}
\end{table*}

\section{Vetting Performance}\label{sec:performance}
\subsection{Flux-Level Vetting}\label{sec:flux}
Using the thresholds provided in \S\ref{sec:noiseFPs} and \S\ref{sec:astroFPs}, \texttt{LEO-Vetter} successfully classified 91.00\% of injected planets as PCs, while rejecting 99.91\% of simulated false alarms as FAs. Only 14 simulated false alarms were not labeled as FAs. Among observed TCEs, we identified $N_{\mathrm{PC}} = 325$, $N_{\mathrm{FP}} = 204$, and $N_{\mathrm{FA}} = 17895$, corresponding to an overall catalog reliability of 96.97\% (i.e., a noise/systematic false alarm rate of 3.03\%) following Eqn. \ref{eqn:reliability}.

\begin{figure*}
    \centering
    \includegraphics[width=0.48\linewidth]{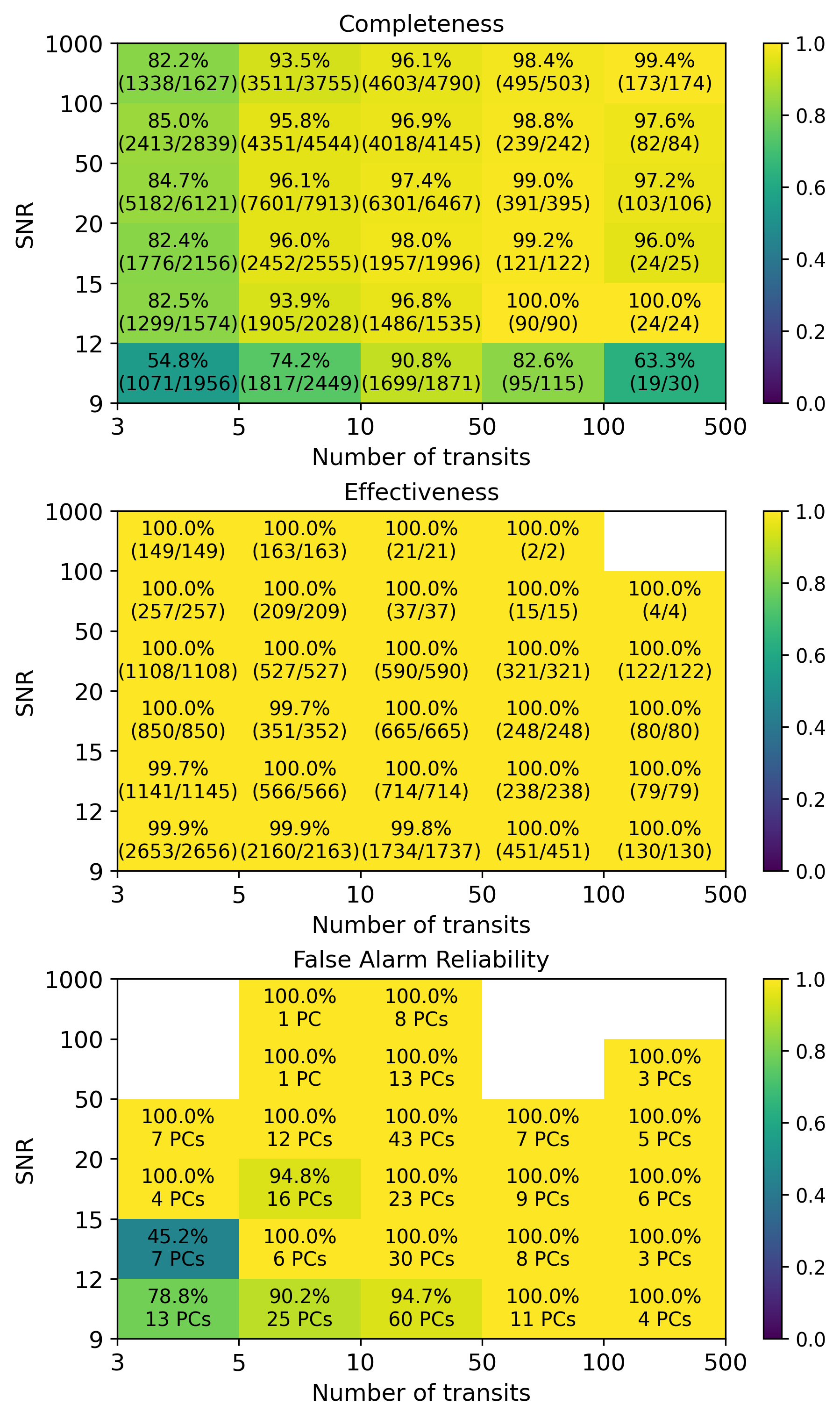}
    \includegraphics[width=0.48\linewidth]{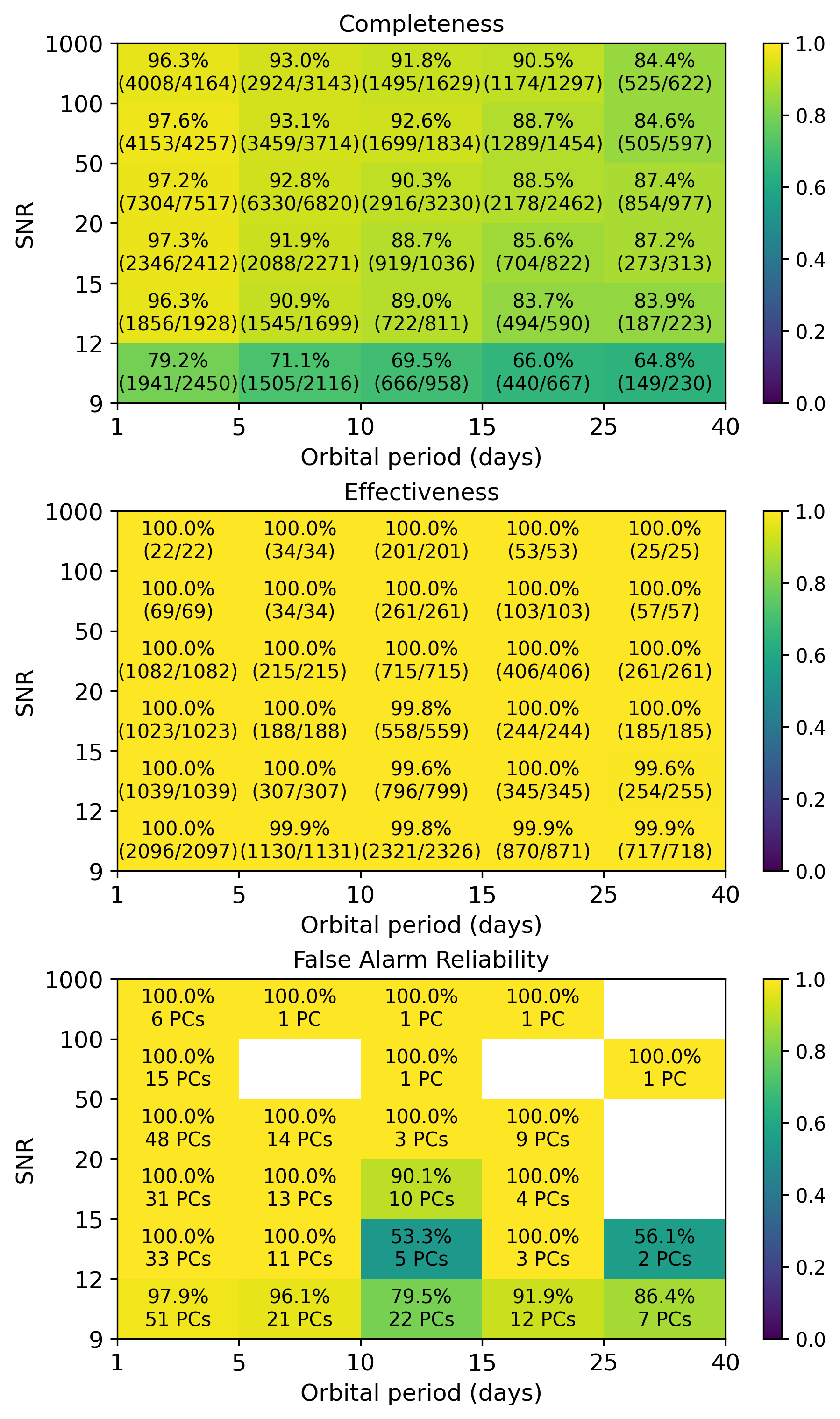}
    \caption{Results from testing \texttt{LEO-Vetter} on simulated planets and false alarms, as functions of SNR and either number of transits (left panels; note that the boundaries between bins are inclusive for the bin to the right, so, e.g., the leftmost boxes have either 3 or 4 candidate transit events) and orbital period (right panels). Because different stars observed by TESS are associated with unique numbers of sectors and severity of data gaps, vetting performance does not directly map between the number of transits and orbital period. Completeness grids report the percentage of injected planets successfully classified as PCs (top numbers) and the actual fraction of injected TCEs classified as PCs (bottom numbers). Effectiveness grids report the percentage of simulated false alarms successfully classified as FAs (top) and the actual fraction of scrambled TCEs classified as FAs (bottom). Reliability grids report the percentage of observed PCs that are expected to be planets (top) and the actual number of PCs identified among observed TCEs (bottom).}\label{fig:results}
\end{figure*}

As shown in Figure \ref{fig:results}, \texttt{LEO-Vetter} performs best for $\mathrm{SNR} > 12$ and $N_{tr} \geq 5$, at which completeness is 96.19\% while reliability is 99.98\%. Completeness generally increases with the number of transits, except for low SNR events with hundreds of transits which are weak, short-period events that are difficult to distinguish from stellar variability. Tests which are only run on TCEs with few events, namely the Chases Test (\S\ref{sec:chases}, only run for $N_{tr} \leq 5$) and the Single Event Domination Test (\S\ref{sec:single}, $N_{tr} \leq 10$), are partially responsible for the drop in completeness at few transits seen in the top-left panel of Figure \ref{fig:results}. Such tests were needed to reduce the enormous number of false alarms detected in the same parameter space (49\% of scrambled TCEs had $N_{tr} \leq 5$, and 67\% of scrambled TCEs had $N_{tr} \leq 10$.) Completeness also generally increases with SNR, except for the highest SNR events which are difficult to distinguish from grazing eclipsing binaries.

The most successful tests for removing false alarms were the Unphysical Transit Duration Tests (\S\ref{sec:unphysical}; 64.32\% of scrambled TCEs removed) and the SNR Consistency Test (\S\ref{sec:CHI}; 63.06\% of scrambled TCEs removed). The former was effective at removing short-period false alarms visible in Figure \ref{fig:tce_hist}, which are primarily caused by stellar variability and contact binary stars. The latter was effective at removing the pileups at one and two times the TESS orbital period, which are primarily caused by scattered light. 

\begin{figure}
    \includegraphics[width=\linewidth]{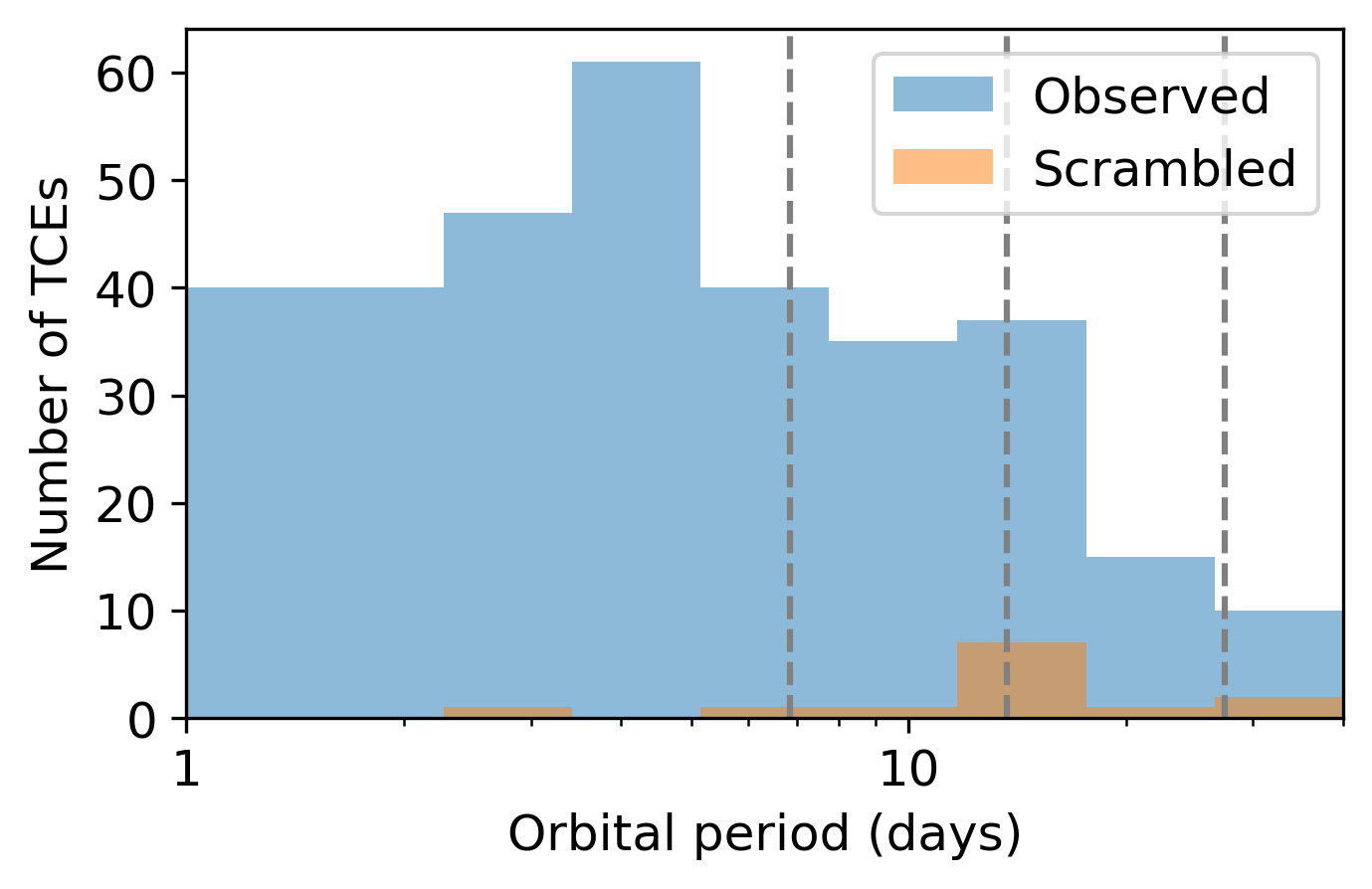}
    \caption{Period histograms from Figure \ref{fig:tce_hist}, but after applying \texttt{LEO-Vetter} flux-level vetting to remove false alarms and false positives and using larger bin sizes due to the smaller number of TCEs. The systematic pileups visible in Figure \ref{fig:tce_hist} are significantly reduced.}\label{fig:pc_hist}
\end{figure}

After applying all flux-level tests (i.e., after removing FAs and FPs), the TCE histogram in Figure \ref{fig:tce_hist} becomes the histogram in Figure \ref{fig:pc_hist}, where short-period TCEs and systematic pileups have been almost entirely removed. The resulting observed TCE histogram has a peak at $P \sim 3$ days and broadly matches the orbital period distribution of TOIs \cite[e.g.,][]{Guerrero2021, Kunimoto2022}.  The period distribution of the surviving scrambled TCEs is also significantly more flat.

Based on ephemeris-matching with the TOI catalog,\footnote{Accessed from ExoFOP on 2024 July 26} our search pipeline detected 147 TOIs as TCEs, of which 129 passed flux-level vetting and 18 failed. Eight of the failed TOIs have already been dispositioned as False Positives by the TESS Follow-Up Observing Program Working Group (TFOPWG), and thus were correctly rejected by \texttt{LEO-Vetter}. Four of the passing TOIs are TFOPWG False Positives, all due to being NEBs. Because flux-level vetting alone is not ideal for flagging NEBs, it is unsurprising that these TOIs passed before pixel-level vetting had been applied. Ignoring these twelve TFOPWG False Positives, flux-level vetting successfully passed of 125 of 135 TOIs. This $\sim93\%$ recovery rate is similar to the $\sim91\%$ completeness estimate based on flux-level vetting performance on injected planets.

TOI-715.01, TOI-4479.01, and TOI-6251.01 are the only TFOPWG Confirmed Planets that failed flux-level vetting. All three TOIs were originally detected in 2-minute cadence data by the TESS Science Processing Operations Center \cite[SPOC;][]{SPOC} and were later statistically validated \citep{Dransfield2024, TOI4479, Kuzuhara2024, Dholakia2024}. We suspect that \texttt{LEO-Vetter} failed these TOIs because the longer-cadence FFI data were not sufficient to distinguish them from low-SNR false alarms, or because the systematics correction applied by SPOC resulted in a cleaner light curve. TOI-715.01 and TOI-4479.01 only barely met our detection criteria ($\mathrm{SNR} = 9.9$ and $9.8$) and failed the SNR Consistency Test with scores of $\mathrm{CHI} = 5.6$ and $5.8$. TOI-6251.01 was also a weak signal ($\mathrm{SNR} = 9.8$) and failed the Chases Test with a mean Chases value of $0.53$. To explore how using shorter cadence data could have improved our results, we re-ran \texttt{LEO-Vetter} using SPOC 2-min cadence light curves in place of the QLP FFI light curves. All three TOIs passed flux-level vetting and would have been considered PCs with the alternative dataset.

\subsection{Performance on Different Cadences}

As mentioned in \S\ref{sec:SES_MES}, \texttt{LEO-Vetter} accounts for different cadences in TESS light curves by weighting each data point according to its uncertainty, given that different exposure times will be associated with different per-point noise. In practice, this affects all \texttt{LEO-Vetter} tests that involve SES and MES values (given that SNR is computed with weighted means and standard deviations), model fitting (given that flux uncertainty is provided as an input to \texttt{lmfit}), and the Rubble test described in \S\ref{sec:indiv} (given that we need to know the cadence of the observations in order to compute the expected number of data points over a stretch of time).

Because different exposure times result in significantly different numbers of data points over the same timespan, we do not expect that \texttt{LEO-Vetter}'s performance will be equivalent across all TESS mission stages. To explore this, we found that 1.7\% of our test set of M dwarfs had only Prime Mission observations (30-minute cadence), 23.6\% had only First Extended Mission observations (10-minute cadence), and 14.9\% had only Second Extended Mission observations (200-second cadence). Each set of light curves have a median observing baseline of one TESS sector. The overall completeness for each mission stage is 69.88\%, 88.25\%, and 89.33\%, respectively, while effectiveness and reliability are 100\% in all three cases. For the other 59.8\% of stars with some mixture of cadences (and therefore more than one sector of observations), overall completeness, effectiveness, and reliability are 92.76\%, 99.87\%, and 95.91\%, respectively. 

These results, in combination with the improved performance on the Confirmed Planet TOIs using 2-minute cadence SPOC observations, demonstrate that \texttt{LEO-Vetter} improves with shorter cadences. We suspect this is due to the increased number of data points available, which improves the vetter's ability to distinguish transit-like events by shape. These results also imply that an alternative solution of re-binning light curves to the same cadence before vetting would result in poorer performance due to the loss of information. 

The most significant improvement in performance comes from the change in cadence between the Prime and First Extended Missions. Users who would like to use \texttt{LEO-Vetter} for, e.g., only Prime Mission observations, will likely want to choose different thresholds to achieve higher completeness.

\subsection{Pixel-Level Vetting}

\texttt{LEO-Vetter} is effective at removing false alarms and false positives based on flux-level vetting, but pixel-level data are needed to reject NEBs, which are significant sources of TESS astrophysical false positives. We ran our Centroid Offset Test (\S\ref{sec:centroid}) on all 325 observed TCEs passing flux-level vetting. We analyzed difference images from the most recent five sectors observed for a given TCE to reduce computational expense. This test failed 153 TCEs, leaving 172 PCs, including 127 TOIs. The adopted threshold of $\Delta\theta < 15^{\prime\prime}$ was sufficient to both remove many NEBs and disposition all planets and planet candidates among TOIs as PCs. Only two TOIs which are known TFOPWG False Positives survived both flux- and pixel-level vetting (TOI-633.01 and TOI-5648.01), both of which are now known to be NEBs, and both of which would have been able to be flagged as false positives using further analysis after \texttt{LEO-Vetter} (\S\ref{sec:PCs}).

\section{Final Uniform Planet Candidate Catalog}\label{sec:PCs}

We have demonstrated that \texttt{LEO-Vetter} is capable of reducing a large number of systematics-dominated TCEs to a significantly smaller set of PCs with both high completeness and high reliability. \texttt{LEO-Vetter} therefore enables additional analyses for planet candidate identification that would otherwise be infeasible to run on all TCEs due to either time or computational expense. These analyses could include more comprehensive transit modeling to improve the accuracy of transit and planet parameters, estimates of astrophysical false positive probabilities with statistical validation packages like \texttt{TRICERATOPS} \citep{Giacalone2021}, more expensive pixel-level tests to identify on- and off-target signals, manual inspection, and observational follow-up.

Some \texttt{LEO-Vetter}-vetted PCs will inevitably still be false alarms or false positives, but following the Kepler occurrence rate literature, we choose to include \textit{all} uniformly vetted candidates in our catalog. Each candidate is supplemented by a set of probabilities to characterize false alarm and astrophysical false positive reliability.

\subsection{Physical and Orbital Parameters}\label{sec:modeling}

\begin{figure*}[t!]
    \centering
    \includegraphics[width=\linewidth]{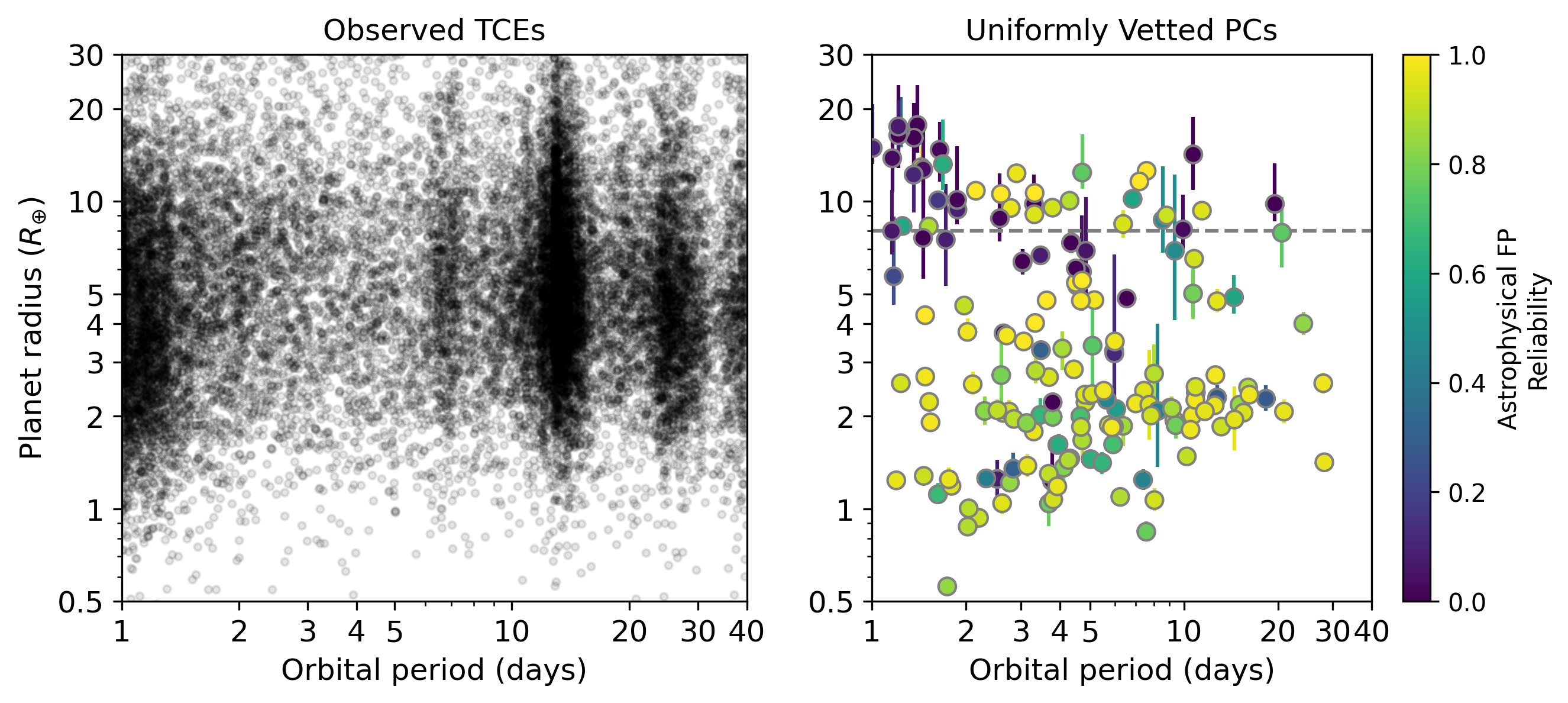}
    \caption{Period-radius diagram of systematics-dominated observed TCEs (left) and only PCs passing both flux- and pixel-level vetting (right). The properties of observed TCEs were estimated using the simple \texttt{LEO-Vetter}-implemented transit model fits, while the properties of PCs were estimated using more comprehensive transit modeling (\S\ref{sec:modeling}). PCs are color-coded by their reliability against astrophysical false positives (\S\ref{sec:reliability}), where higher values mean that the candidate is more likely to be a real planet. However, reliability will not be robust for planets larger than $R_{p} = 8~R_{\oplus}$ (dotted grey line) due to degeneracies with brown dwarfs and low-mass stars \citep{Giacalone2021}.}\label{fig:per-rad}
\end{figure*}

We followed the general procedure described by \citet{Rowe2015} and \citet{Rowe2016} to determine the final physical and orbital parameters of the 172 PCs in our uniformly vetted catalog.  Modeling was done with a Keplerian orbit and a transit model incorporating quadratic limb-darkening \citep{MandelAgol2002}.  The lightcurve was parameterized by the mean-stellar density ($\bar \rho_{\star}$), quadratic limb-darkening ($q_1$, $q_2$) \citep{Kipping2016}, orbital period ($P$), center-of-transit time ($T_0$), scaled planetary radius ($R_{p} / R_{\star}$) and impact parameter ($b$).  We assumed a circular orbit for the presented models, as the relatively low SNR of the discovered transits is not sensitive to the photo-eccentricity effect \citep{Dawson2012}.  We adopted Gaussian priors on $q_1$ and $q_2$ based on stellar parameters reported in Table \ref{tab:stars} by generating quadratic fits to interpolated grids from the MPS-ATLAS library \citep{Kostogryz2023} integrated over the TESS band pass based on the mean and standard deviation of 1000 random samples of T$_{\rm eff}$, $\log{g}$ and [Fe/H] with a cut-off radius of $\mu$ = 0.1.   The MCMC posterior modeling, updated from that described in \citet{Rowe2016}, was set to sample in $\log{\bar \rho_{\star}}$ and $\log{ R_{p} / R_{\star} }$ to avoid {\it pseudo-density} bias \citep{Gilbert2022}.  MCMC sampling used a difference MCMC technique as described in \citet{Gregory2011}.  Chains were evaluated for convergence based on having Gelman-Rubin metric less than 1.02 for all model parameters.  Chains have a minimum length of 200,000.  Posteriors of model parameters were parameterized by calculating the mode and the 68\% credible interval from a kernel-density-estimator, except for $b$, where we report the median value (see \S2.5 of \citealt{Lissauer2014} for details).  These parameters are provided in Table \ref{tab:PCs} along with host star properties in Table \ref{tab:stars}, and a period-radius diagram is shown in Figure \ref{fig:per-rad}.

\begin{sidewaystable}
    \centering
    \begin{tabular}{ccc|ccccccc|c}
\hline\hline
TIC & TOI & CTOI & $P$ & $T_{0}$ & $R_{p}/R_{\star}$ & $b$ & $\bar \rho_{\star}$ & $q_{1}$ & $q_{2}$ & $R_{p}$ \\
 & & & (days) &(BJD - 2457000) & & & & & & ($R_{\oplus}$)\\
\hline
4487172 & - & - &$24.12953_{-0.0004000}^{+0.0003800}$ & $1853.315_{-0.01300}^{+0.01400}$ & $0.071_{-0.005}^{+0.006}$ & $0.390_{-0.360}^{+0.200}$ & $2.60_{-1.40}^{+1.20}$ & $0.375_{-0.014}^{+0.013}$ & $0.222_{-0.017}^{+0.017}$ & $4.00_{-0.32}^{+0.36}$ \\
4619242 & 5295.01 & - &$4.302482_{-0.0000058}^{+0.0000065}$ & $2450.2971_{-0.00090}^{+0.00089}$ & $0.158_{-0.005}^{+0.004}$ & $0.692_{-0.054}^{+0.074}$ & $4.27_{-0.97}^{+1.41}$ & $0.395_{-0.015}^{+0.016}$ & $0.202_{-0.020}^{+0.019}$ & $10.04_{-0.49}^{+0.43}$ \\
7439480 & 6110.01 & - &$1.25412353_{-0.0000007}^{+0.0000008}$ & $1600.30583_{-0.00059}^{+0.00071}$ & $0.139_{-0.002}^{+0.004}$ & $0.430_{-0.150}^{+0.210}$ & $7.98_{-3.10}^{+0.31}$ & $0.383_{-0.014}^{+0.014}$ & $0.212_{-0.019}^{+0.017}$ & $8.31_{-0.32}^{+0.30}$ \\
12421862 & 198.01 & - &$10.215207_{-0.0000260}^{+0.0000210}$ & $1356.3724_{-0.00330}^{+0.00300}$ & $0.030_{-0.002}^{+0.002}$ & $0.460_{-0.230}^{+0.450}$ & $6.00_{-3.90}^{+1.50}$ & $0.367_{-0.012}^{+0.013}$ & $0.236_{-0.015}^{+0.017}$ & $1.48_{-0.09}^{+0.11}$ \\
12999193 & - & 12999193.01 &$1.48264123_{-0.0000003}^{+0.0000004}$ & $1355.44431_{-0.00043}^{+0.00033}$ & $0.155_{-0.004}^{+0.004}$ & $0.300_{-0.270}^{+0.150}$ & $453.00_{-120.00}^{+70.00}$ & $0.381_{-0.017}^{+0.016}$ & $0.185_{-0.023}^{+0.018}$ & $4.26_{-0.18}^{+0.15}$ \\
16005254 & 5344.01 & - &$3.79281_{-0.0002100}^{+0.0001800}$ & $2477.3118_{-0.00110}^{+0.00150}$ & $0.151_{-0.008}^{+0.004}$ & $0.650_{-0.110}^{+0.140}$ & $4.60_{-1.50}^{+3.50}$ & $0.384_{-0.015}^{+0.014}$ & $0.211_{-0.017}^{+0.018}$ & $9.53_{-0.54}^{+0.41}$ \\
19028197 & 5094.01 & - &$3.336661_{-0.0000230}^{+0.0000220}$ & $2503.87203_{-0.00027}^{+0.00031}$ & $0.073_{-0.001}^{+0.001}$ & $0.260_{-0.230}^{+0.130}$ & $5.25_{-0.99}^{+0.21}$ & $0.385_{-0.015}^{+0.014}$ & $0.217_{-0.019}^{+0.015}$ & $4.02_{-0.13}^{+0.12}$ \\
20182780 & 3984.01 & - &$4.3533295_{-0.0000107}^{+0.0000099}$ & $1931.42422_{-0.00097}^{+0.00110}$ & $0.143_{-0.004}^{+0.007}$ & $0.590_{-0.160}^{+0.210}$ & $2.50_{-0.91}^{+2.63}$ & $0.397_{-0.017}^{+0.014}$ & $0.197_{-0.015}^{+0.023}$ & $7.35_{-0.34}^{+0.39}$ \\
22233480 & 4438.01 & - &$7.446177_{-0.0000350}^{+0.0000540}$ & $2396.4133_{-0.00180}^{+0.00160}$ & $0.059_{-0.002}^{+0.002}$ & $0.360_{-0.330}^{+0.180}$ & $9.00_{-3.40}^{+1.30}$ & $0.396_{-0.016}^{+0.014}$ & $0.200_{-0.019}^{+0.019}$ & $2.42_{-0.10}^{+0.13}$ \\
23863106 & 5648.01 & - &$3.30807_{-0.0006400}^{+0.0006800}$ & $2672.5212_{-0.00190}^{+0.00240}$ & $0.175_{-0.025}^{+0.040}$ & $0.760_{-0.200}^{+0.240}$ & $1.99_{-0.90}^{+2.77}$ & $0.384_{-0.015}^{+0.014}$ & $0.211_{-0.017}^{+0.018}$ & $9.80_{-1.40}^{+2.40}$ \\
... & ... & ... & ... & ... & ... & ... & ... & ... & ... & ... \\
\end{tabular}
    \caption{Transit model properties of the 172 PCs passing flux- and pixel-level vetting. The full table is available in a machine-readable format.}\label{tab:PCs}
\end{sidewaystable}

\begin{table*}
    \centering
    \begin{tabular}{c|cccc}
\hline\hline
TIC & $T$ (mag) & $R_{\star}~(R_{\odot})$ & $M_{\star}~(M_{\odot})$ & $T_{\mathrm{eff}}$ (K) \\
\hline
4487172 & 13.36 &$0.52\pm0.02$ &$0.52\pm0.02$ &$3715\pm157$ \\
4619242 & 13.02 &$0.58\pm0.02$ &$0.57\pm0.02$ &$3504\pm157$ \\
7439480 & 13.07 &$0.54\pm0.02$ &$0.54\pm0.02$ &$3642\pm157$ \\
12421862 & 9.93 &$0.45\pm0.01$ &$0.44\pm0.02$ &$3782\pm157$ \\
12999193 & 13.37 &$0.25\pm0.01$ &$0.22\pm0.02$ &$3478\pm157$ \\
16005254 & 13.22 &$0.58\pm0.02$ &$0.57\pm0.02$ &$3575\pm157$ \\
19028197 & 10.25 &$0.5\pm0.01$ &$0.5\pm0.02$ &$3552\pm157$ \\
20182780 & 13.46 &$0.46\pm0.01$ &$0.46\pm0.02$ &$3422\pm157$ \\
22233480 & 11.27 &$0.38\pm0.01$ &$0.36\pm0.02$ &$3353\pm157$ \\
23863106 & 13.9 &$0.51\pm0.02$ &$0.51\pm0.02$ &$3571\pm157$ \\
... & ... & ... & ... & ... \\
\end{tabular}
    \caption{Host star properties of the 172 PCs passing flux- and pixel-level vetting. The full table is available in machine-readable format.}\label{tab:stars}
\end{table*}

\subsection{Astrophysical False Positive Reliability}\label{sec:reliability}

As demonstrated in \S\ref{sec:flux}, \texttt{LEO-Vetter} can be used to characterize reliability against false alarms on a population level. Some TCEs may alternatively be astrophysical false positives, namely on- or off-target eclipsing binaries. Knowledge of reliability against both false alarms and false positives is essential for robust occurrence rate calculations \cite[e.g.,][]{Bryson2020}.

The statistical validation package \texttt{TRICERATOPS} provides an estimate of FPP (False Positive Probability), reflecting the probability that an observed event is due to an astrophysical false positive around the target star \citep{Giacalone2021}. One method of measuring astrophysical false positive reliability is to simply set $R_{\mathrm{FP}} = 1 - \mathrm{FPP}$. As part of the FPP calculation, \texttt{TRICERATOPS} also computes NFPP$_{i}$ (Nearby False Positive Probability), reflecting the probability that the observed transit originates from the $i$-th nearby resolved star rather than the target star, based on the light curve and magnitudes of nearby stars. The total probability that the event is off-target is $\mathrm{NFPP} = \sum_{i=1}\mathrm{NFPP}_{i}$. We include both FPP and NFPP values for all 172 PCs in Table \ref{tab:probs}. Note, however, that FPP values from statistical validation techniques alone are not robust for $R_{p} > 8~R_{\oplus}$ due to degeneracies with brown dwarfs and low-mass stars \citep{Giacalone2021}; spectroscopic follow-up is needed to confirm the true natures of these events.\footnote{Dynamical criteria can serve as an alternative to spectroscopic follow-up for some of candidate giant planets in multi-planet systems; see Appendix D of \cite{Lissauer2014} for details, but are not used for the analysis presented herein.}

\begin{sidewaystable}
    \centering
    \begin{tabular}{ccc|cccccccc}
\hline\hline
TIC & TOI & CTOI & Comment & SNR & FPP & NFPP & APP$_{0}$ & Astrophysical False & False Alarm & Disposition \\
 & & & & & & & & Positive Reliability & Reliability & Score \\
\hline
4487172 & - & - & - & $9.56$ & $0.007$ & $0.000$ & $0.835$ & $0.993$ & $0.902$ & $0.467$ \\
4619242 & 5295.01 & - & TFOPWG APC & $27.61$ & $0.111$ & $0.000$ & $1.000$ & $0.889$ & $1.000$ & $0.999$ \\
7439480 & 6110.01 & - & TFOPWG PC & $25.32$ & $0.078$ & $0.046$ & $0.618$ & $0.939$ & $1.000$ & $1.000$ \\
12421862 & 198.01 & - & TFOPWG CP & $14.59$ & $0.015$ & $0.000$ & $0.918$ & $0.985$ & $0.948$ & $0.766$ \\
12999193 & - & 12999193.01 & Best Candidate & $30.18$ & $0.002$ & $0.000$ & $1.000$ & $0.998$ & $1.000$ & $0.999$ \\
16005254 & 5344.01 & - & TFOPWG PC & $28.70$ & $0.071$ & $0.000$ & $1.000$ & $0.929$ & $1.000$ & $0.991$ \\
19028197 & 5094.01 & - & TFOPWG KP & $135.45$ & $0.000$ & $0.000$ & $1.000$ & $1.000$ & $1.000$ & $0.999$ \\
20182780 & 3984.01 & - & TFOPWG CP & $34.07$ & $0.093$ & $0.000$ & $0.356$ & $0.907$ & $1.000$ & $0.970$ \\
22233480 & 4438.01 & - & TFOPWG CP & $23.52$ & $0.047$ & $0.000$ & $0.999$ & $0.953$ & $1.000$ & $0.992$ \\
23863106 & 5648.01 & - & TFOPWG FP & $13.84$ & $0.622$ & $0.277$ & $0.015$ & $0.020$ & $1.000$ & $1.000$ \\
... & ... & ... & ... & ... & ... & ... & ... & ... & ... \\
\end{tabular}
    \caption{False positive probabilities, reliabilities, and disposition scores for all PCs. Astrophysical false positives reliabilities are based on \texttt{TRICERATOPS} and pixel-level positional probability analysis, while false alarm reliabilities are based on the coarse grid shown in Figure \ref{fig:results} by SNR and number of transits. Astrophysical false positive reliabilities together with false alarm reliabilities close to 1.0 indicate strong planet candidacy. Of the 172 PCs, 127 are known TOIs, and have TFOPWG Dispositions indicating whether they are currently considered a known planet (KP), confirmed planet (CP), planet candidate (PC), ambiguous planet candidate (APC), or false positive (FP). Manually selected best new planet candidates are indicated with the comment ``Best Candidate.'' The full table is available in a machine-readable format.}\label{tab:probs}
\end{sidewaystable}

An improved estimate of reliability can be obtained by including pixel-level information. For each resolved star considered to be a possible transit host by \texttt{TRICERATOPS}, we estimate APP$_{i}$ (Astrophysical Positional Probability), the probability that the observed event is on the $i$-th star, through the centroid-free Bayesian probability analysis of pixel-level data from \cite{Bryson2025}. This method forms difference images for each sector as described in \cite{Bryson2013}, compares them with simulated difference images that assume the transit is on each nearby star, and provides a probability, APP$_{i}$ for each star.  APP$_{0}$ refers to the probability that the event is on the target star. 

The likelihood that an observed event is caused by a transiting planet on the target star is $(1 - \mathrm{FPP})\times \mathrm{APP}_{0}$, while the likelihood that an event is on the $i$-th nearby star is $\mathrm{NFPP}_{i}\times\mathrm{APP}_{i}$. The total astrophysical false positive reliability will therefore be
\begin{equation}
    R_{\mathrm{FP}} = \frac{(1 - \mathrm{FPP})\times\mathrm{APP}_{0}}{(1 - \mathrm{NFPP})\times\mathrm{APP}_{0} + \sum_{i=1}(\mathrm{NFPP}_{i}\times\mathrm{APP}_{i})}.
\end{equation}

The APP and reliability values are included in Table \ref{tab:probs}. For comparison, we also include the TFOPWG Dispositions of all TOIs. Two PCs are TFOPWG False Positives (TOI-633.01 and TOI-5648.01). The pixel-level analysis showed that TOI-633.01 has only a $0.6\%$ probability of being on-target and a 99.4\% probability of being on the nearby star TIC-348755728 ($15.7^{\prime\prime}$ away), which is the same star that TFOP determined was the correct source. The pixel-level analysis also showed that TOI-5648.01 has only a $1.5\%$ probability of being on-target and a 98.5\% probability of being on the nearby TIC-23863105 ($7.0^{\prime\prime}$ away), which also matches the TFOP follow-up conclusions. These false positives have very low reliabilities ($R_{\mathrm{FP}} = 0.0004$ and $0.020$, respectively).

\subsection{Disposition Scores}\label{sec:dispScore}

Following \cite{Thompson2018}, we compute a false-alarm disposition score for the PCs in our catalog.  This disposition score indicates the confidence with which \texttt{LEO-Vetter} determines that the entries in our catalog are not false alarms. This is accomplished by measuring the standard deviation of \texttt{LEO-Vetter} FA metrics as a function of period and SNR, and measuring how often a PC passes the FA metric tests when considering that standard deviation.  Because the disposition score does not consider the rate of observed false alarms, this disposition score is not the same as reliability.  The relationship between the disposition score and reliability for Kepler data is examined by \cite{Bryson2020}.

\begin{figure}
    \includegraphics[width=\linewidth]{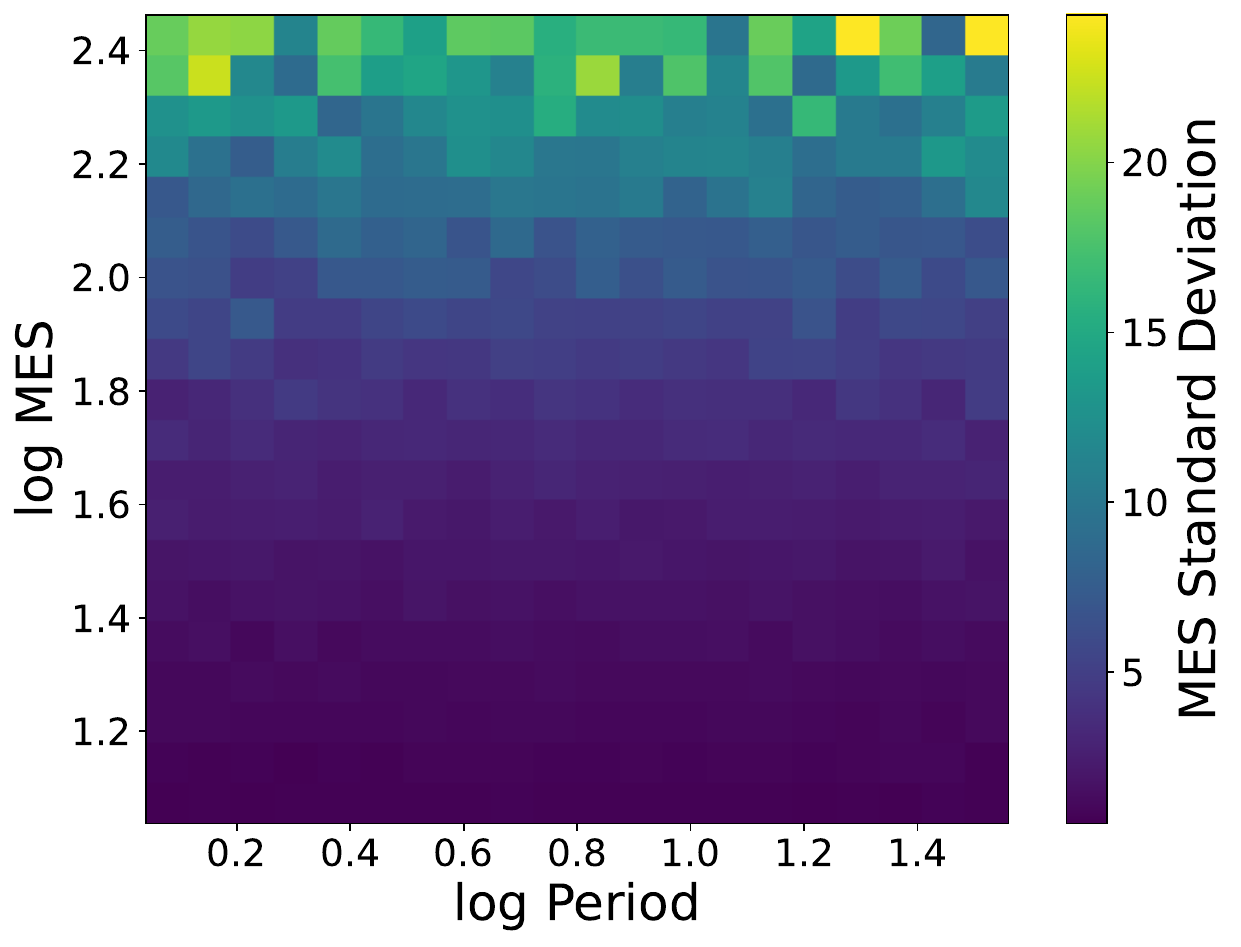}
    \caption{An example of the analysis described in \S\ref{sec:dispScore}, demonstrating the standard deviation of the MES metric as a function of period and MES.  There is a similar grid for every metric.}\label{fig:mes}
\end{figure}

We use injected light curves to measure the standard deviation of the \texttt{LEO-Vetter} metrics as a function of period and SNR for simulated ``true'' transit signals.  For each false alarm metric, we place the injected metric value on the period-SNR plane in the domain $1 < \mathrm{period} < 40$ days and $7.7 < \mathrm{SNR} < 320$.  This gives us 60,694 injected transit signals.  The period-SNR plane is divided into $30 \times 30$ bins evenly spaced in $\log(\mathrm{period})$ and $\log(\mathrm{SNR})$.  The median number of entries in a bin is 55, with a maximum of 211 entries.  All but one bin contain five or more injected signals, and that one bin contains one signal.  We treat each bin as a distribution of metric values for which we measure the standard deviation for that bin.  For most metrics, we measure the standard deviation as the median absolute deviation divided by 0.67449, although for the Chases metric we measure the standard deviation directly because most values of Chases are 0. An example 2D grid for the standard deviation of the MES metric is shown in Figure \ref{fig:mes}.

We observe that for all FA metrics the distribution of standard deviations shows a smooth dependence on period and SNR.  To create a function that can be evaluated at a planet candidate's period and SNR, we fit polynomials to the standard deviation bins.  We consider three types of polynomials: 1D polynomials in period, 1D polynomials in SNR, and 2D polynomials in both period and SNR.  For each type, the order of the polynomials is chosen to minimize the AIC.  We then select the polynomial type with the lowest AIC.  For all metrics, the residual of the best polynomial fit is close to Gaussian. 

The disposition score for each PC is computed by first getting a set of metric standard deviations by evaluating the metric standard deviation polynomials at that planet's period and SNR.  Then the \texttt{LEO-Vetter} FA tests are run 1000 times, where each time the metric thresholds are drawn from Gaussian distributions, with means set to the optimized threshold values and standard deviations set to the metric standard deviations for this PC.  The disposition score for this PC is the fraction of times this PC passes these 1000 tests.  PCs with metrics far from the thresholds will have a high fraction, and PCs with metrics closer to the thresholds will pass less often.  As expected, candidates with lower SNR and higher period have lower disposition scores.

\subsection{Most Promising New Planet Candidates}
Based on visual inspection of the 45 new PCs (non-TOIs), none appeared to be obvious false alarms that should have failed automated flux-level vetting, though seven appeared to be off-target based on their difference images. Contamination from a small number of NEBs in the PC catalog is unsurprising given that we adopted a relatively lenient $\Delta\theta < 15^{\prime\prime}$ threshold for the Centroid Offset Test, which was empirically chosen based on the performance of the test on the TOI catalog. All seven of these PCs have APP$_{0} < 0.1$ based on the positional probability analysis and are low-reliability PCs.

Of the remaining 38 PCs, we identified 21 as our most promising new candidates (see Table \ref{tab:PCs} and Figure \ref{fig:best}). The rest were deemed either too low-SNR and difficult to distinguish from false alarms, or too large and grazing ($R_p > 10~R_{\oplus}$, $b > 0.9$), and therefore are likely EBs. Three of the 21 best signals are already known Community TESS Objects of Interest (CTOIs) detected by \citet{Eschen2024} (TIC-12999193.01, 262605715.01, 311276853.01), three are CTOIs detected by \citet{Montalto2020} (TIC-111778581.01, 153919886.01, 154809195.01), one is a CTOI detected by \citet{Feliz2021} (TIC-290048573.01), and one is a confirmed planet \cite[TIC-46432937 b;][]{Hartman2024}. The other 13 are new to this work.

\begin{figure*}
\centering
    \includegraphics[width=0.31\linewidth]{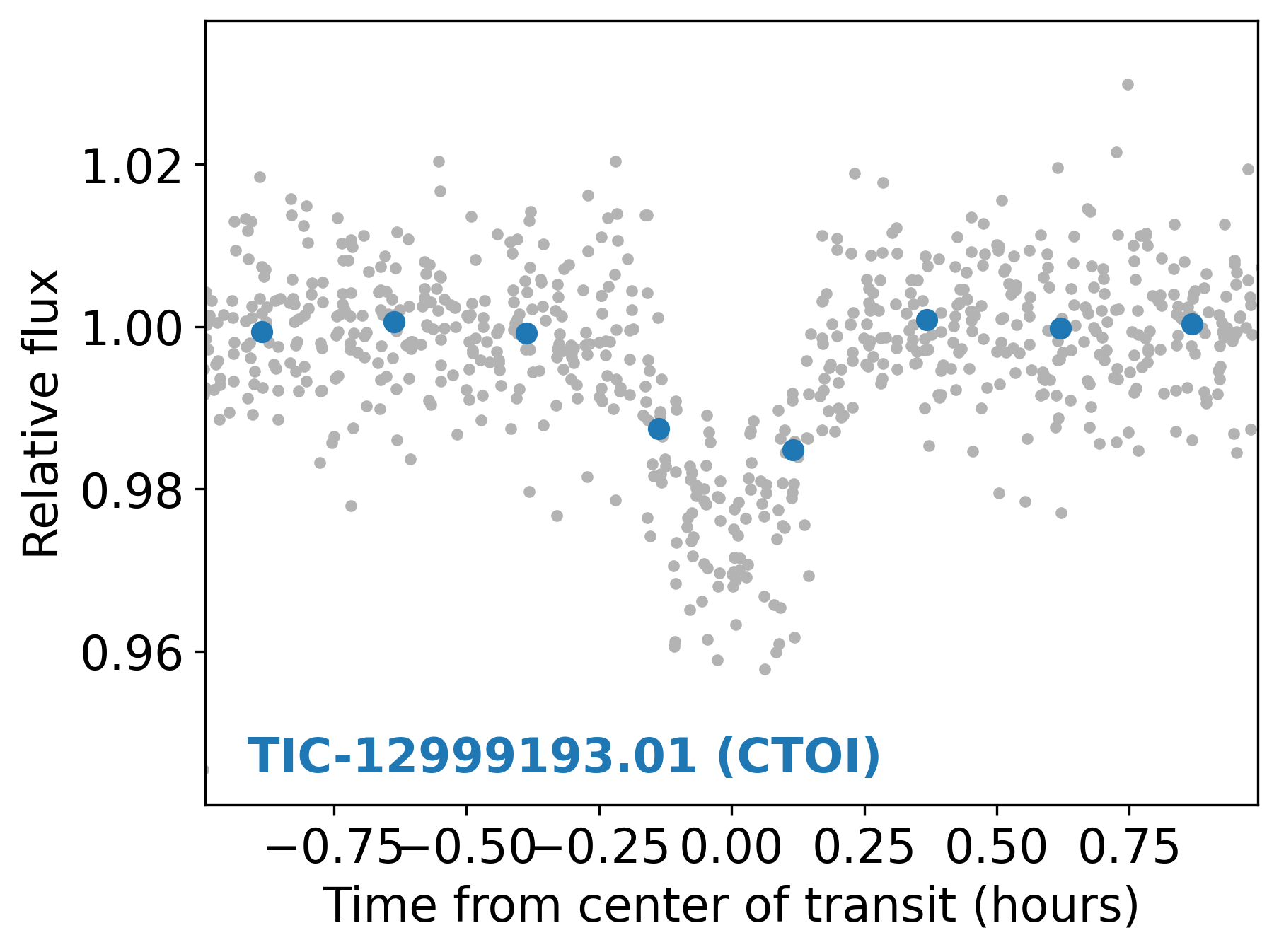}
    \includegraphics[width=0.31\linewidth]{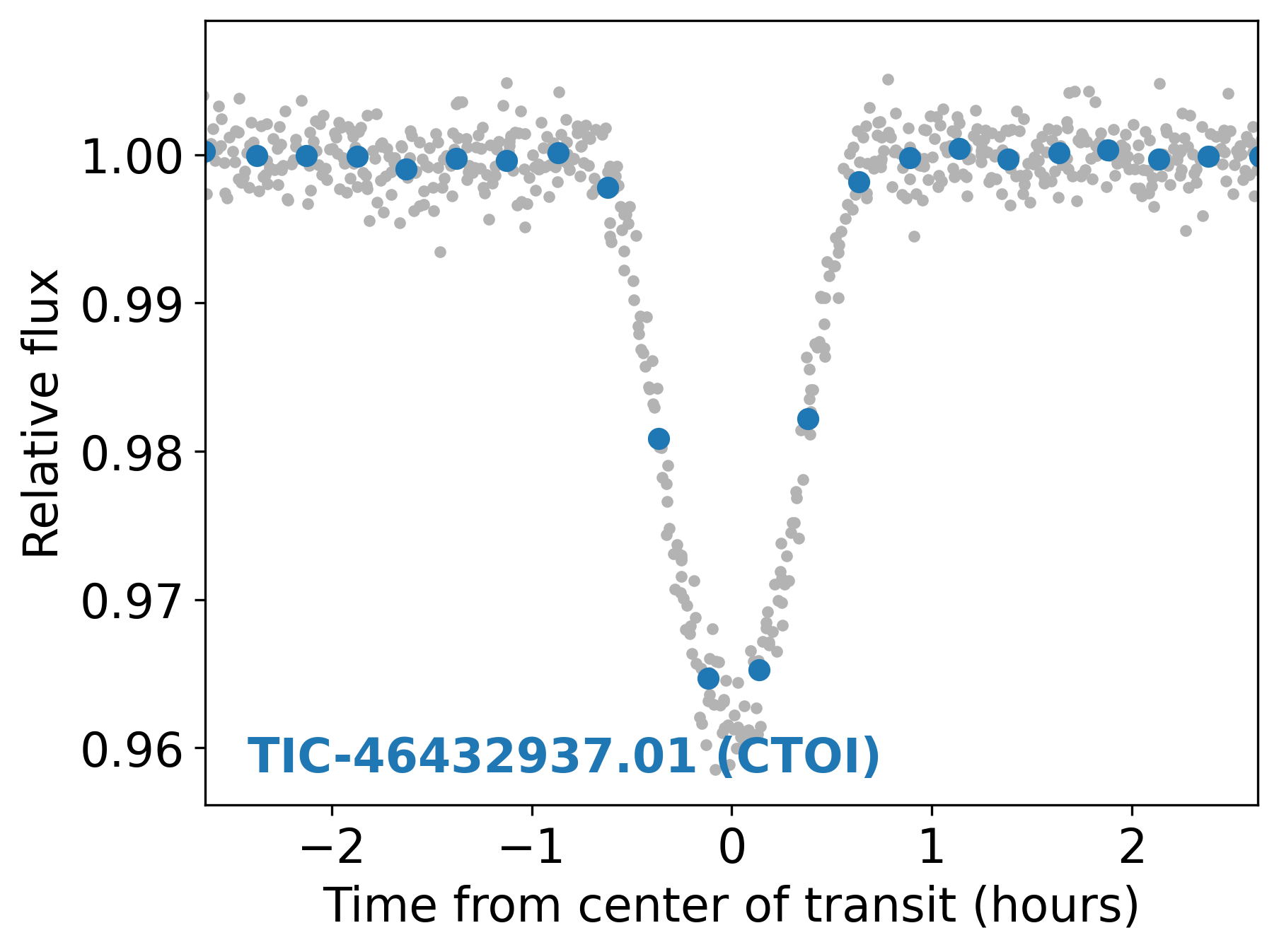}
    \includegraphics[width=0.31\linewidth]{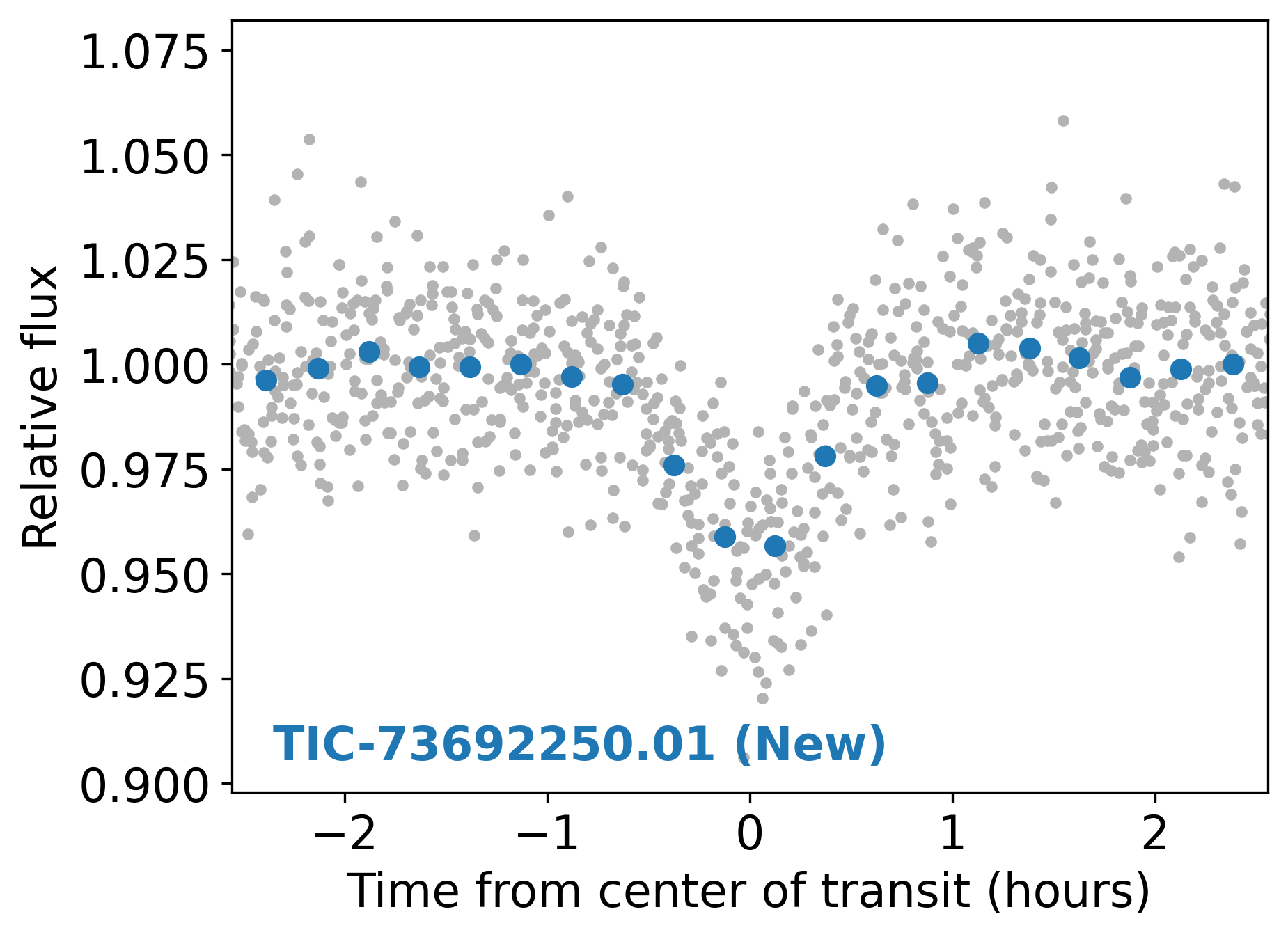}
    \includegraphics[width=0.31\linewidth]{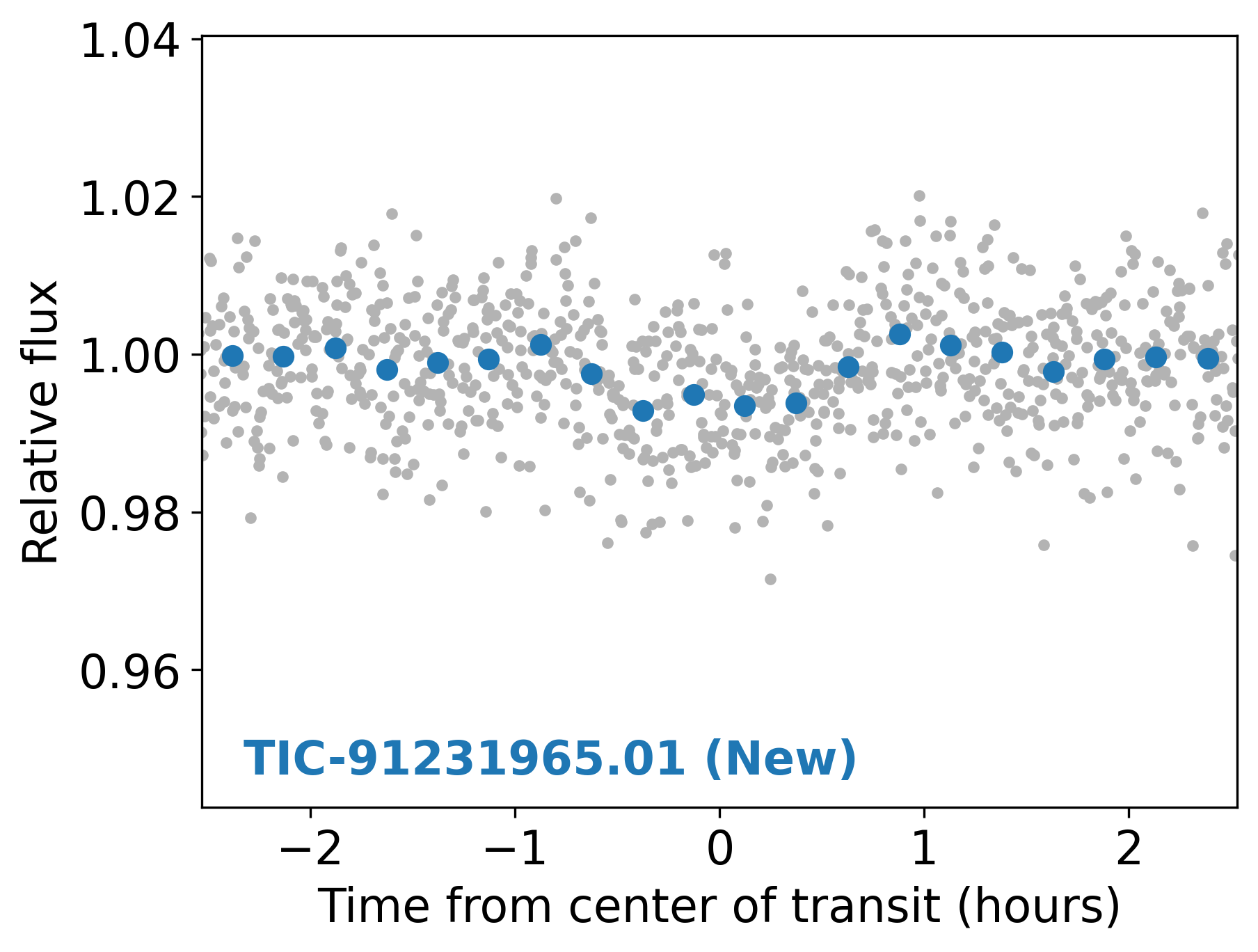}
    \includegraphics[width=0.31\linewidth]{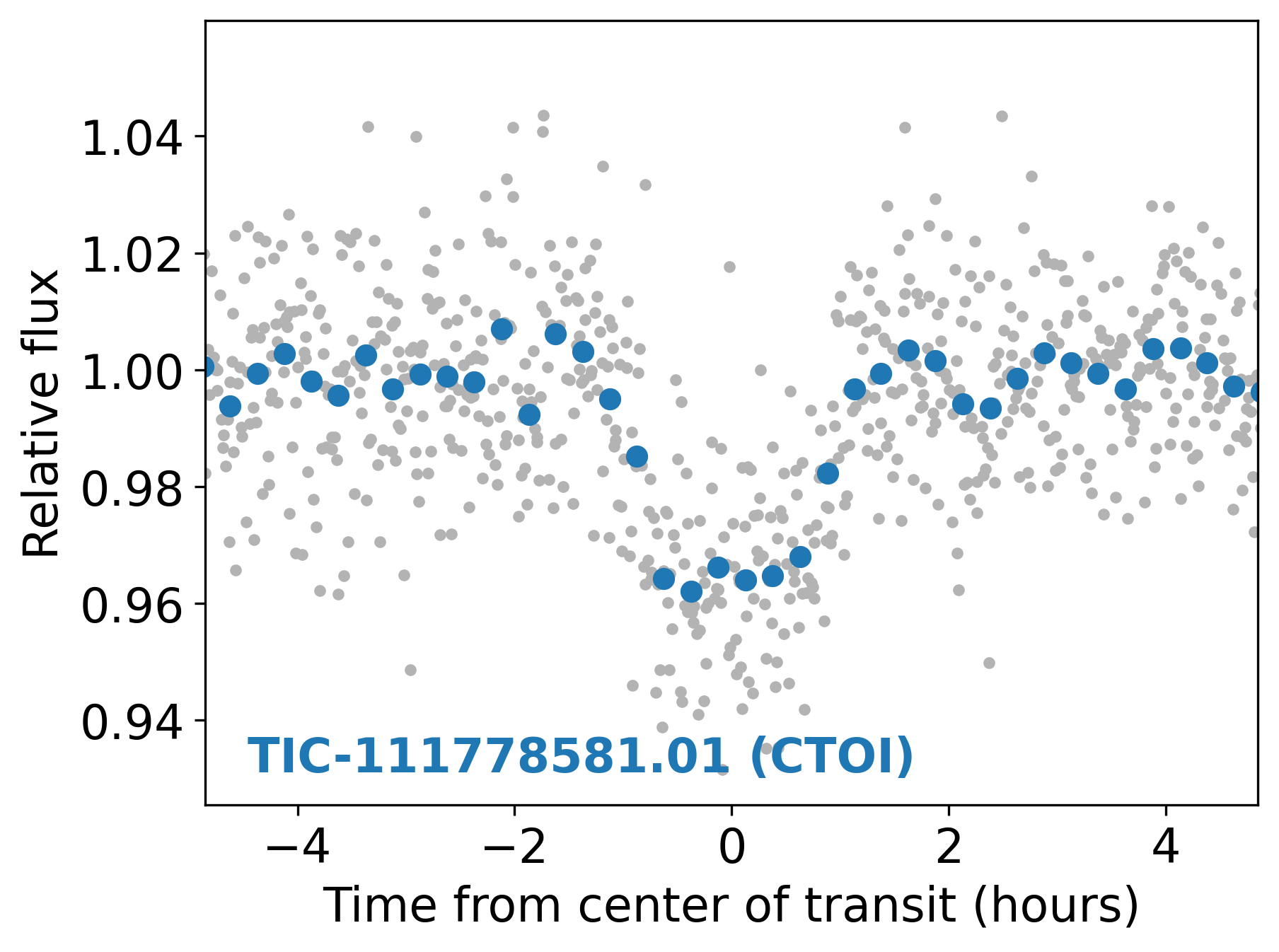}
    \includegraphics[width=0.31\linewidth]{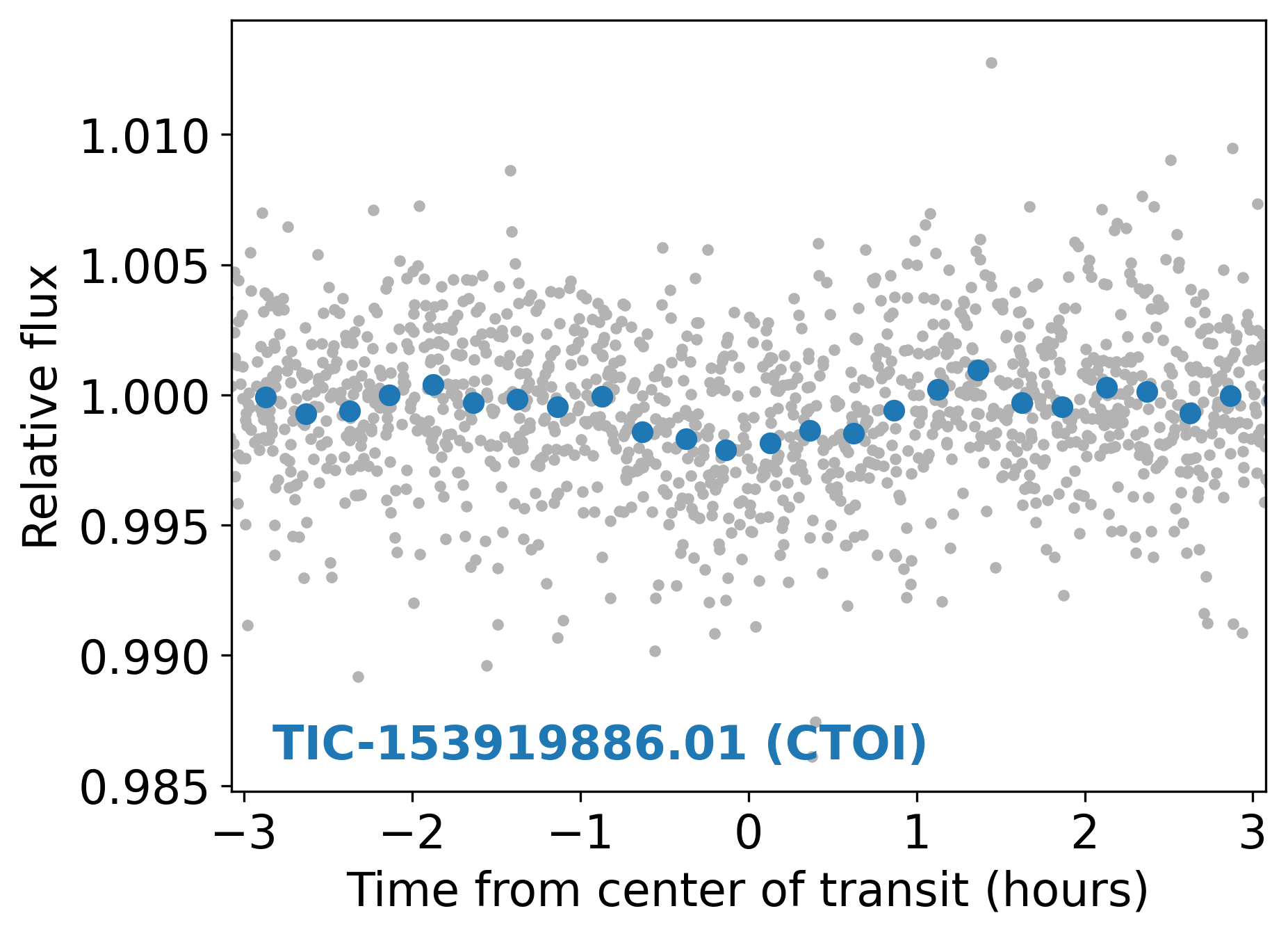}
    \includegraphics[width=0.31\linewidth]{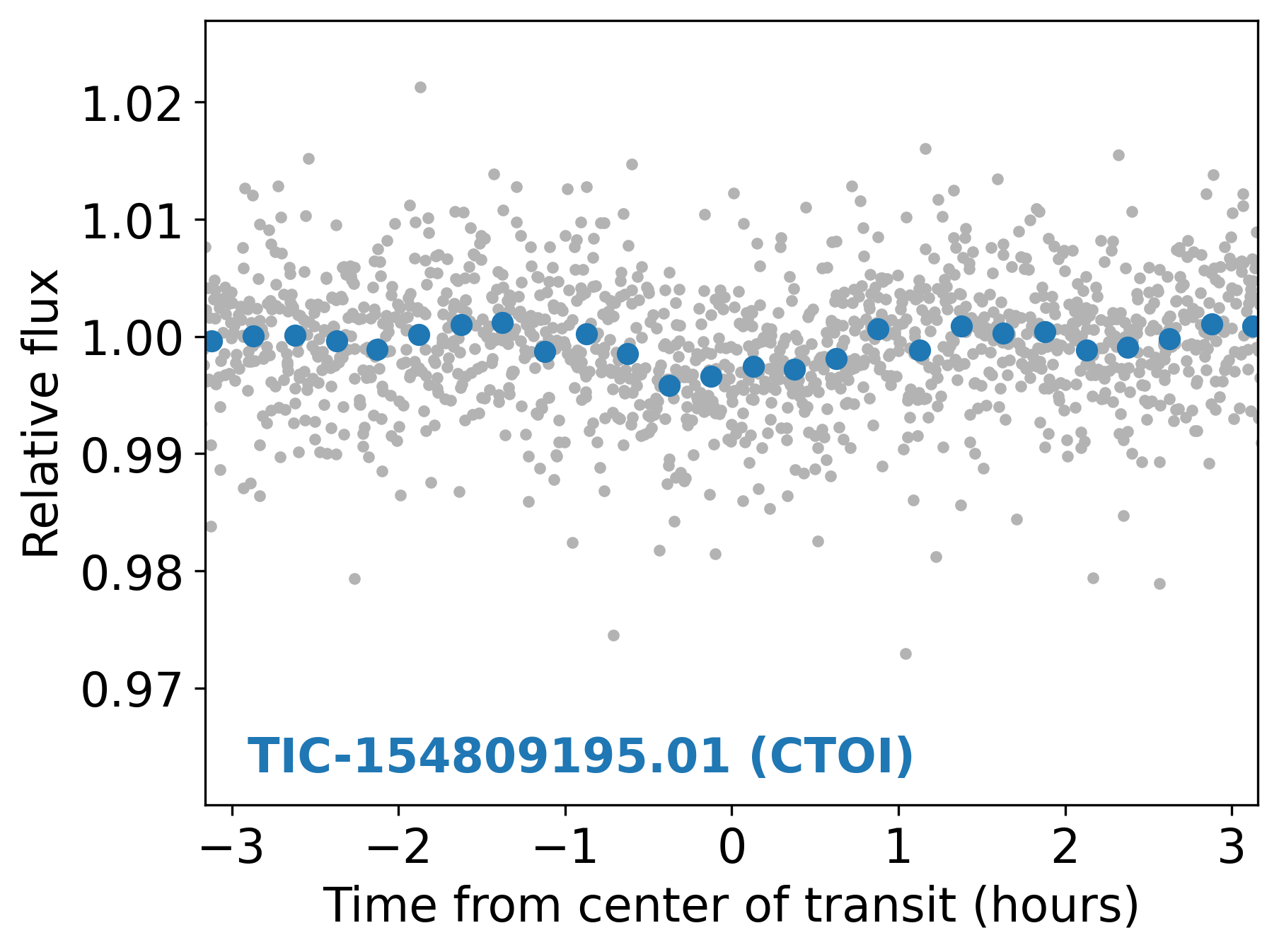}
    \includegraphics[width=0.31\linewidth]{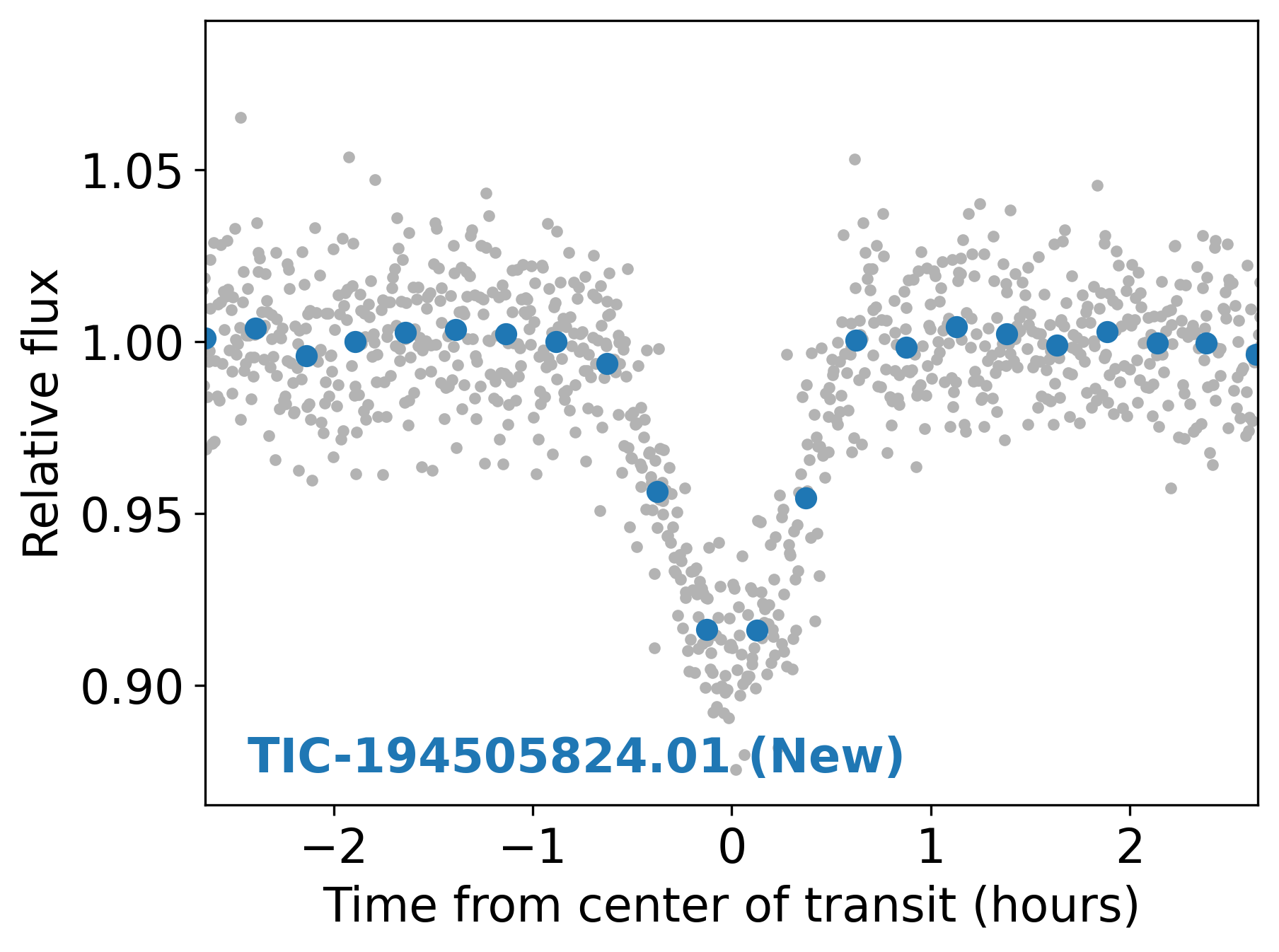}
    \includegraphics[width=0.31\linewidth]{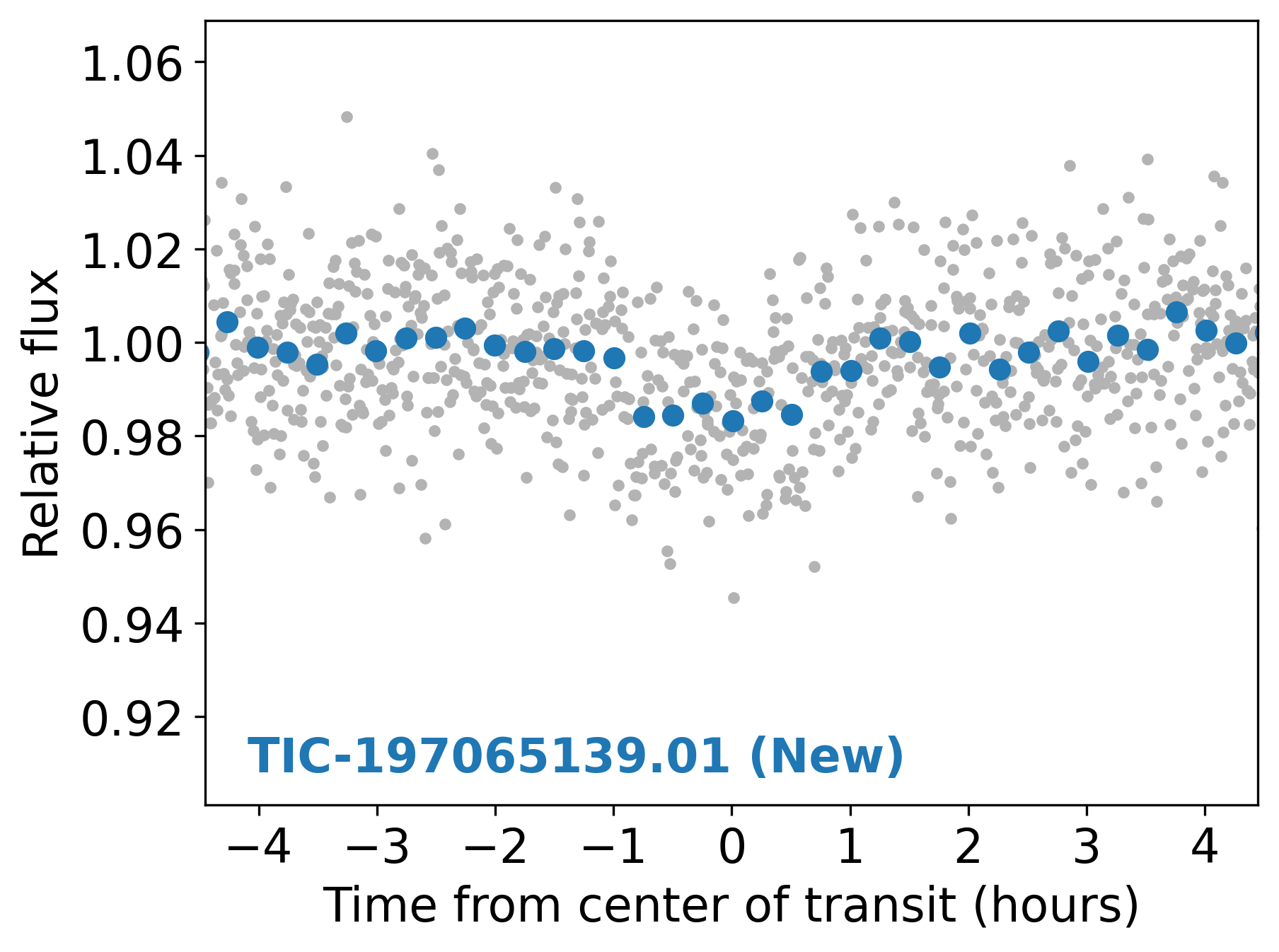}
    \includegraphics[width=0.31\linewidth]{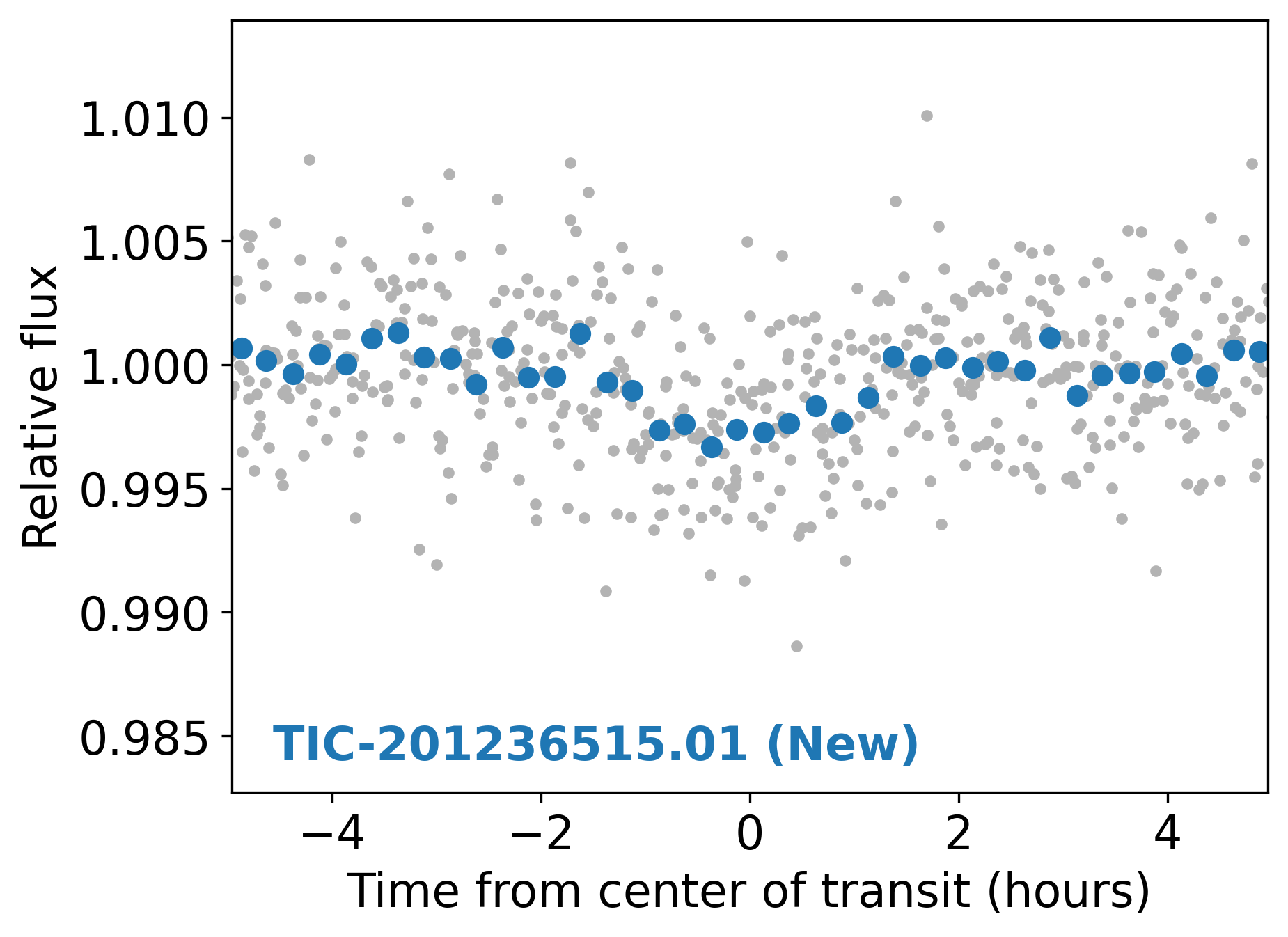}
    \includegraphics[width=0.31\linewidth]{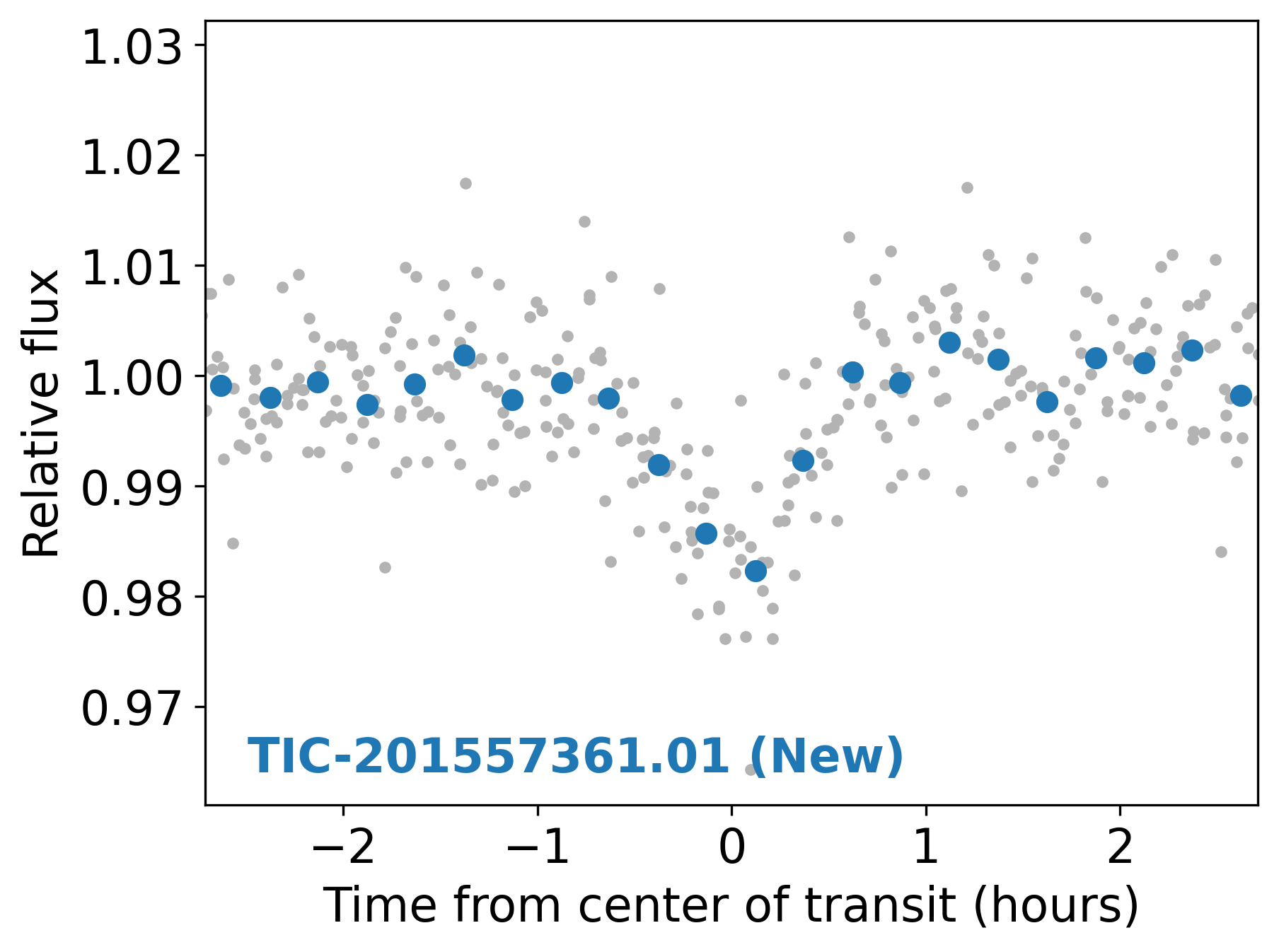}
    \includegraphics[width=0.31\linewidth]{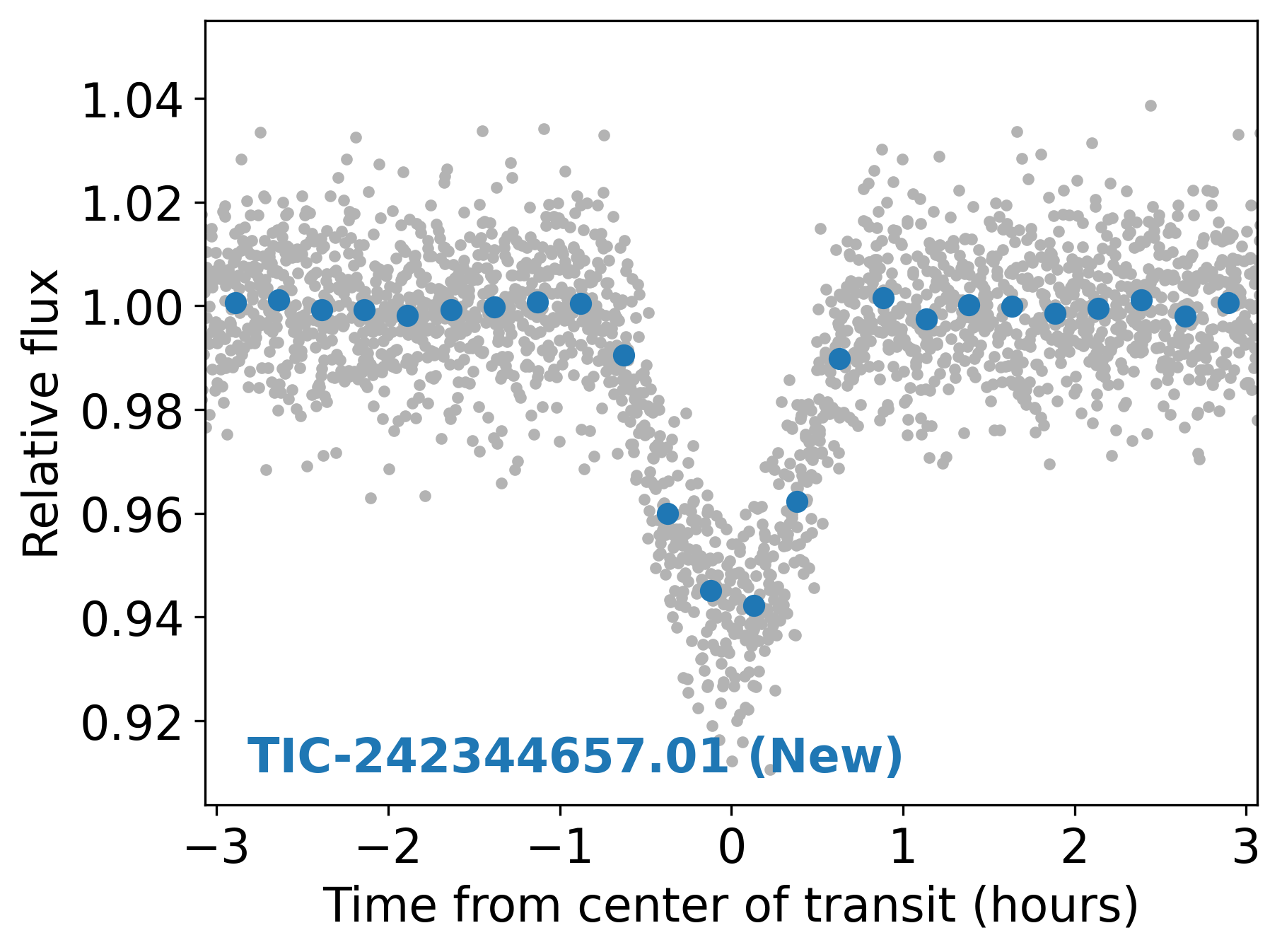}
    \includegraphics[width=0.31\linewidth]{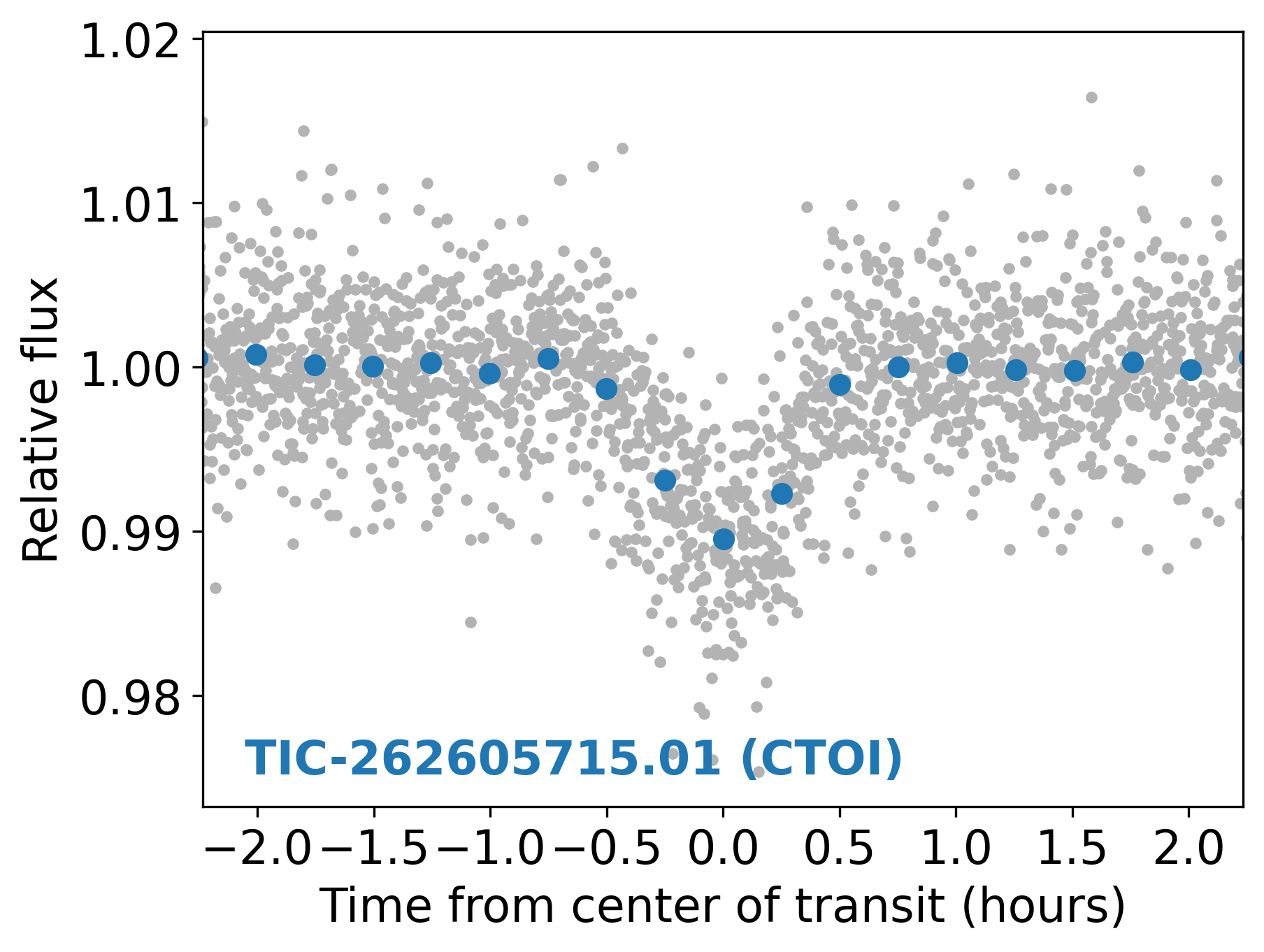}
    \includegraphics[width=0.31\linewidth]{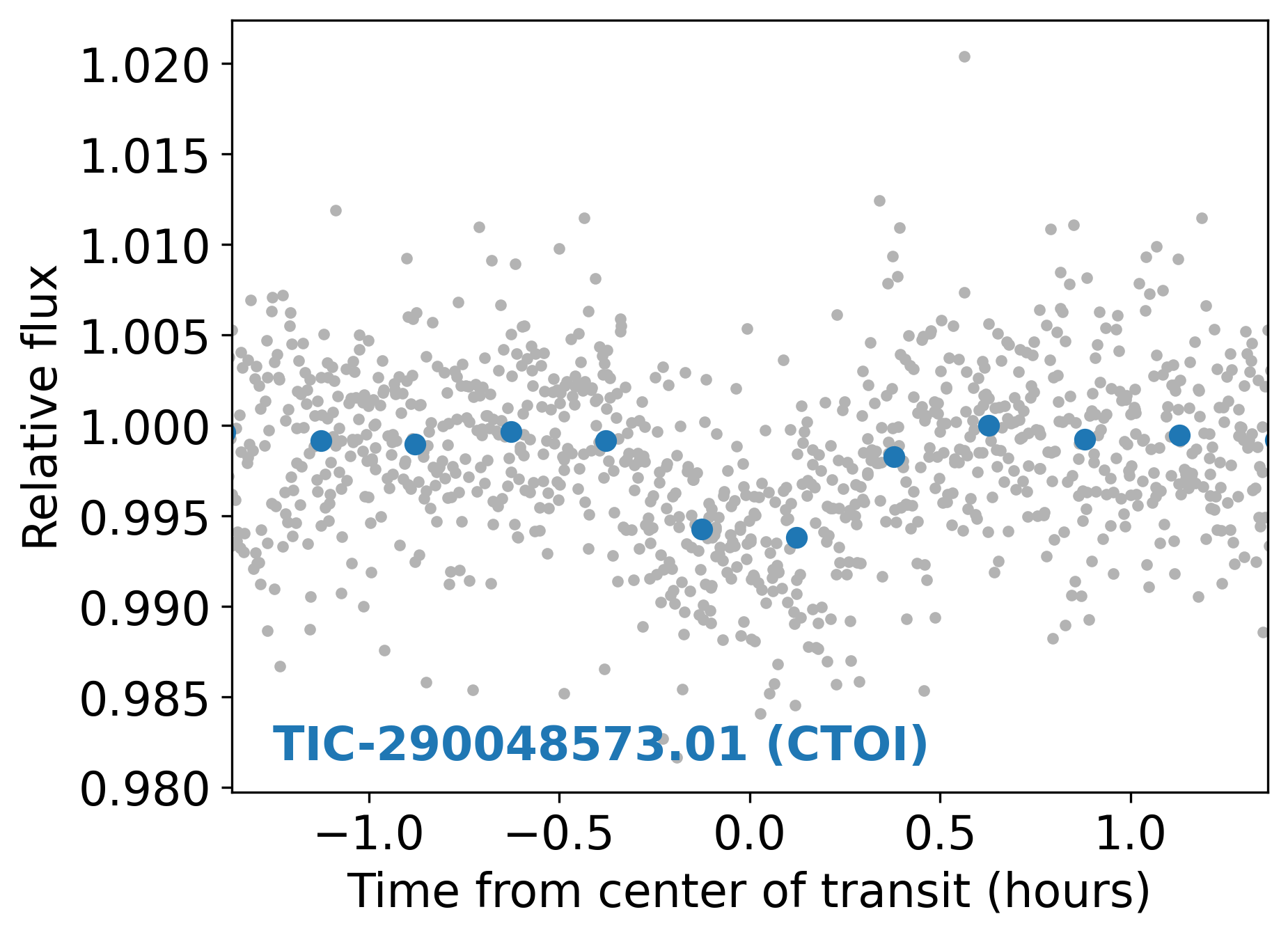}
    \includegraphics[width=0.31\linewidth]{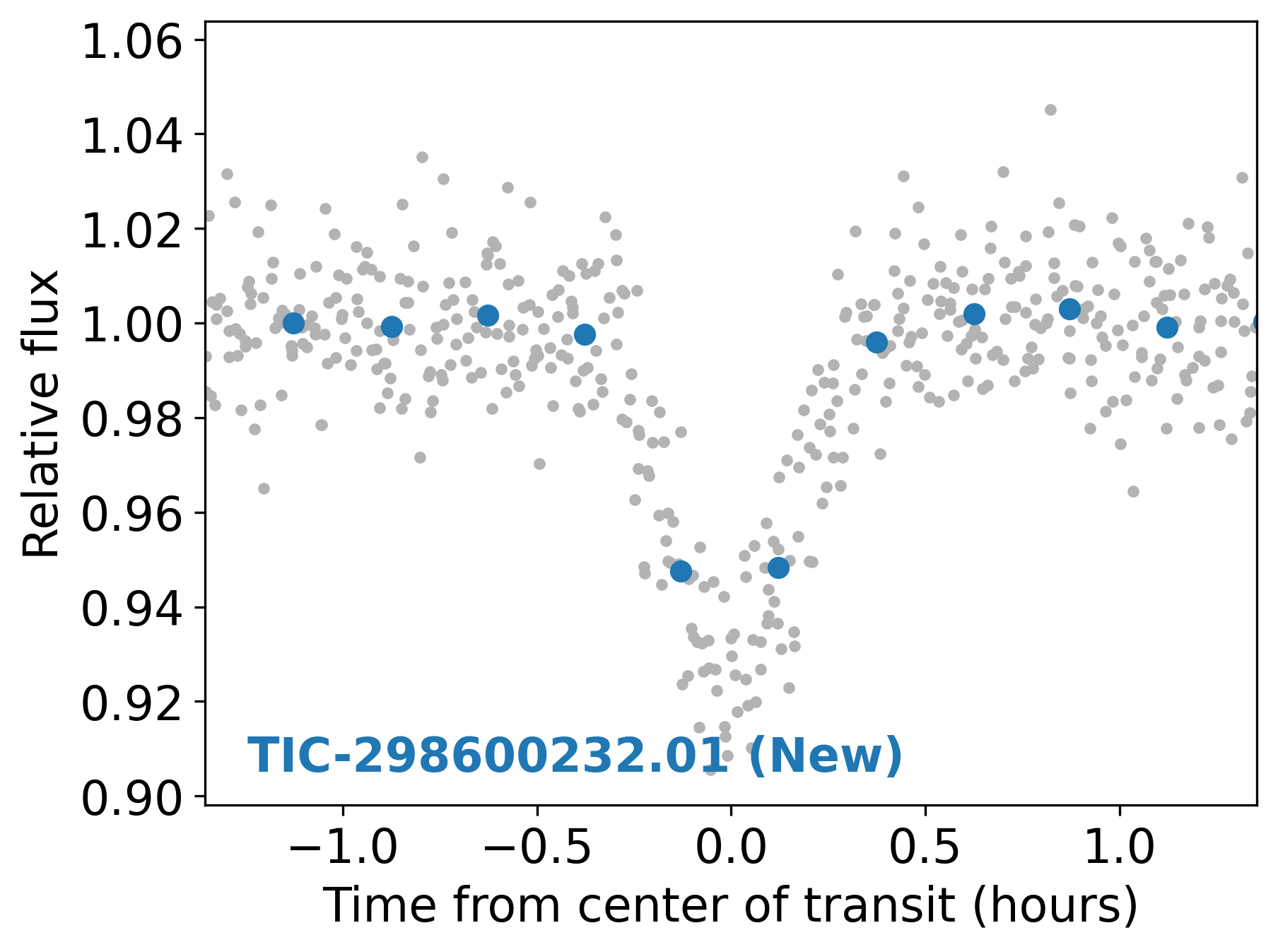}
    \caption{Promising PCs identified by \texttt{LEO-Vetter} that are not yet TOIs. The thirteen PCs which are new CTOIs as a result of this work are indicated by the label ``New''. The rest are known CTOIs from other works \citep{Eschen2024,Montalto2020,Feliz2021,Hartman2024}. Grey dots show the underlying TESS data, while blue dots show the data binned every 15 minutes.}\label{fig:best}
\end{figure*}

\setcounter{figure}{7}

\begin{figure*}[t!]
\centering
    \includegraphics[width=0.31\linewidth]{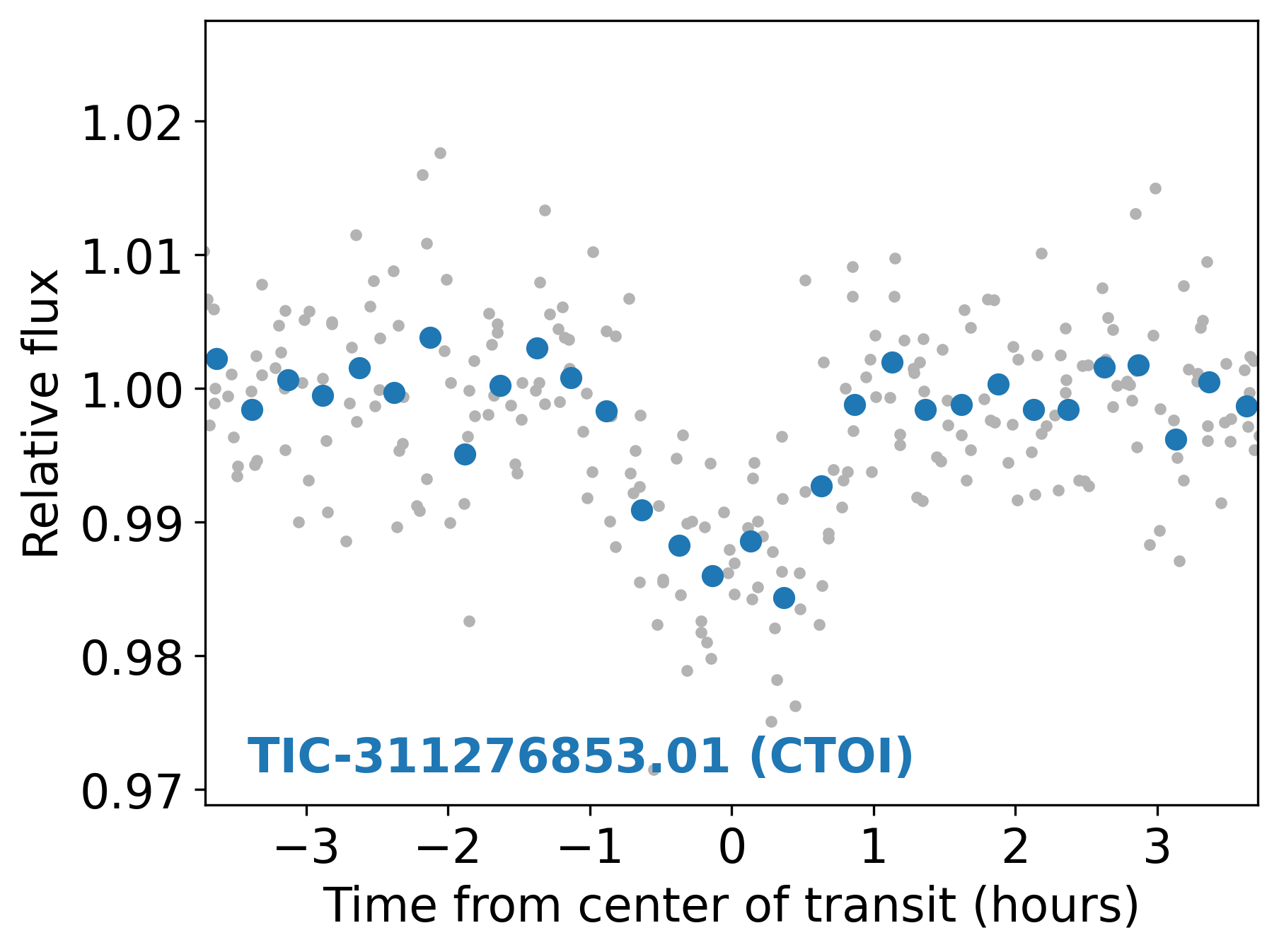}
    \includegraphics[width=0.31\linewidth]{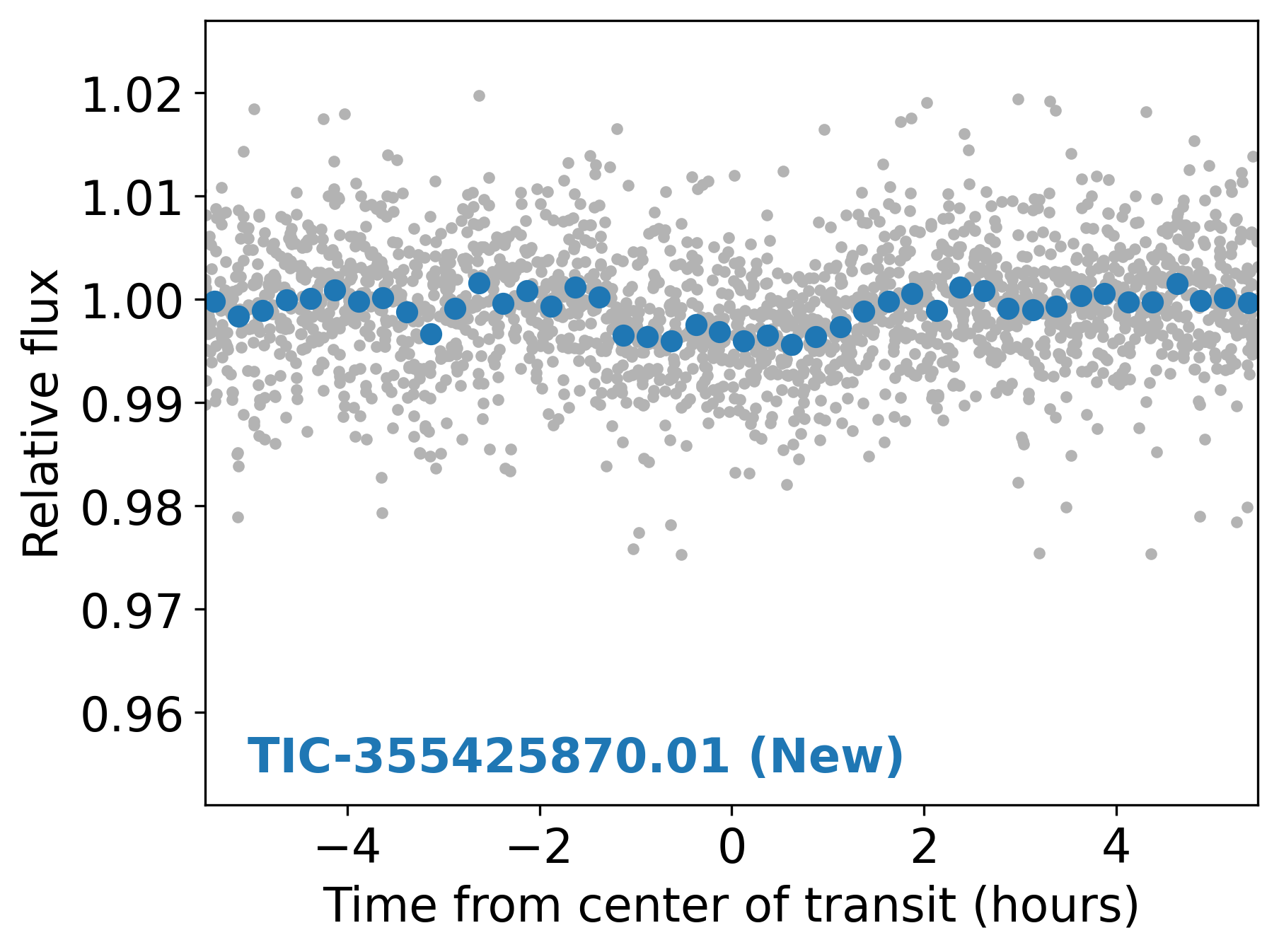}
    \includegraphics[width=0.31\linewidth]{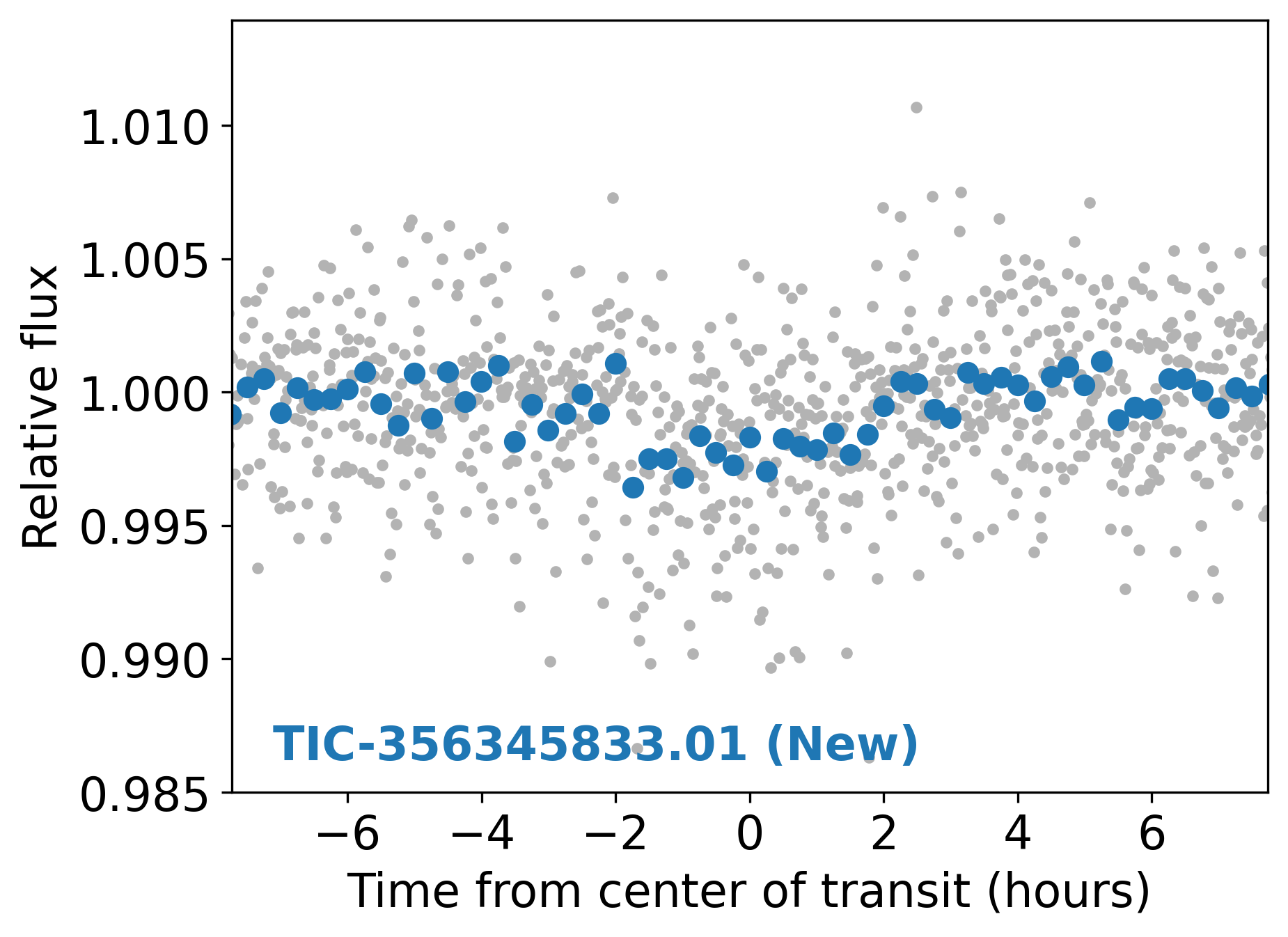}
    \includegraphics[width=0.31\linewidth]{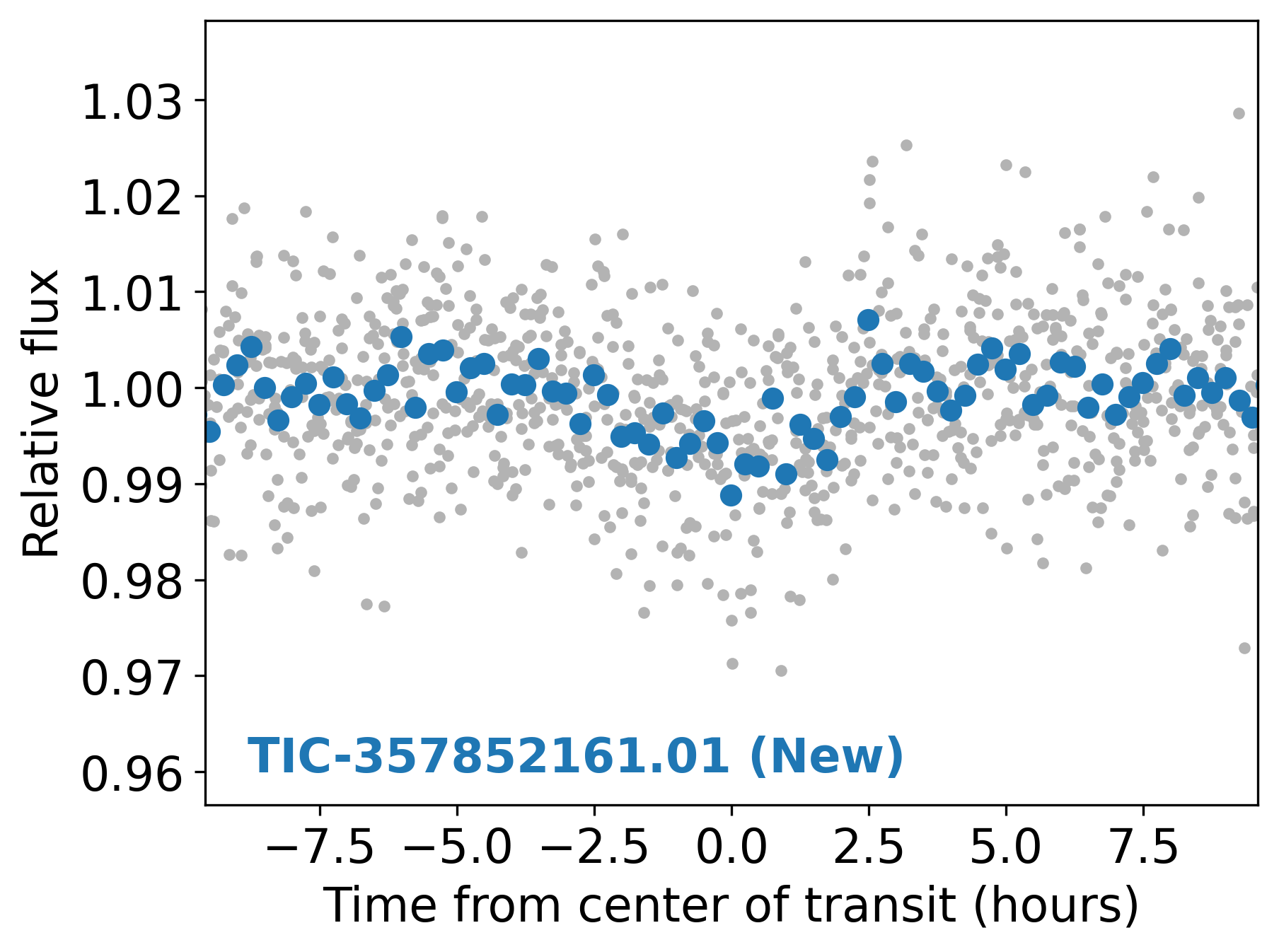}
    \includegraphics[width=0.31\linewidth]{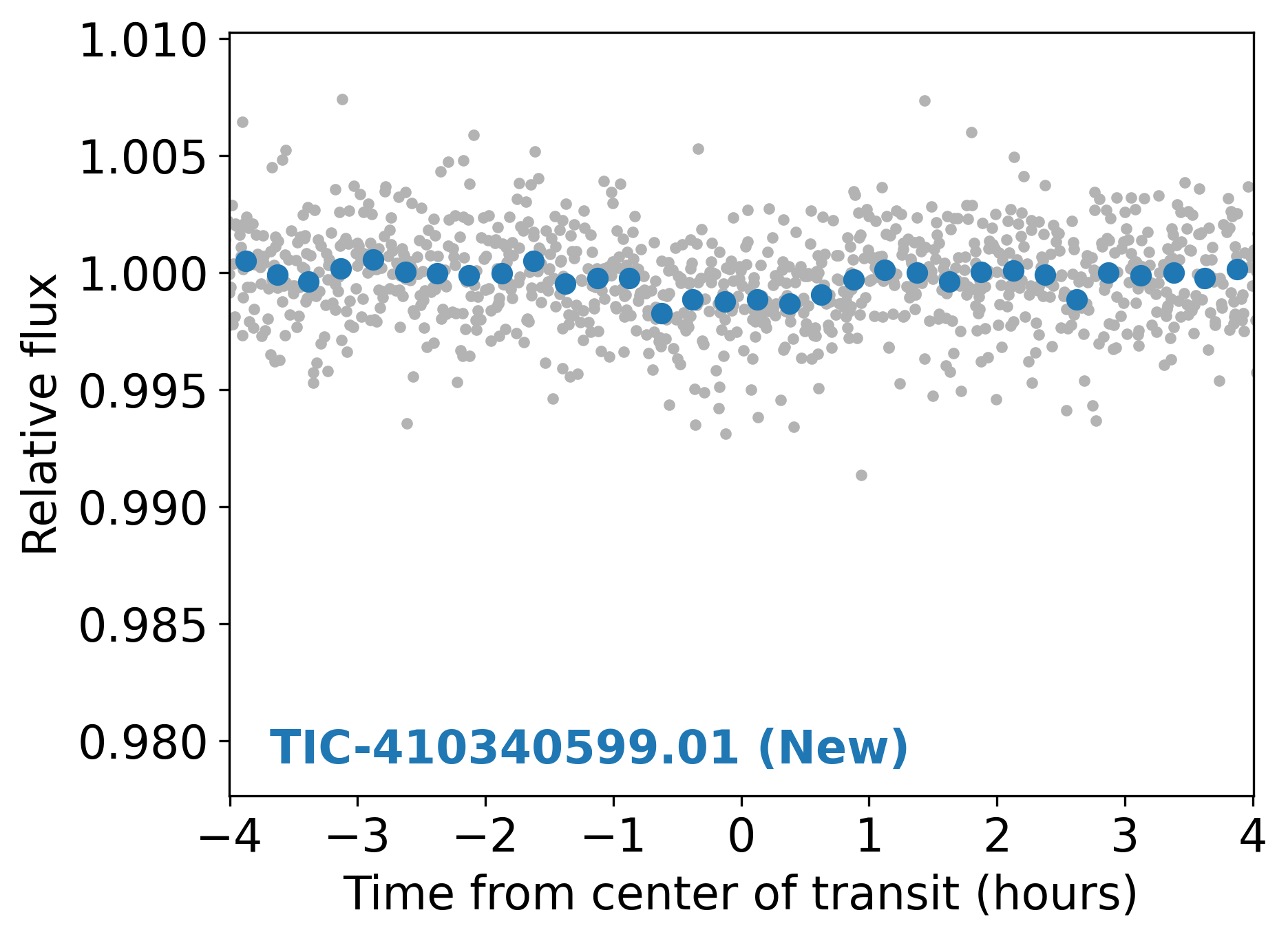}
    \includegraphics[width=0.31\linewidth]{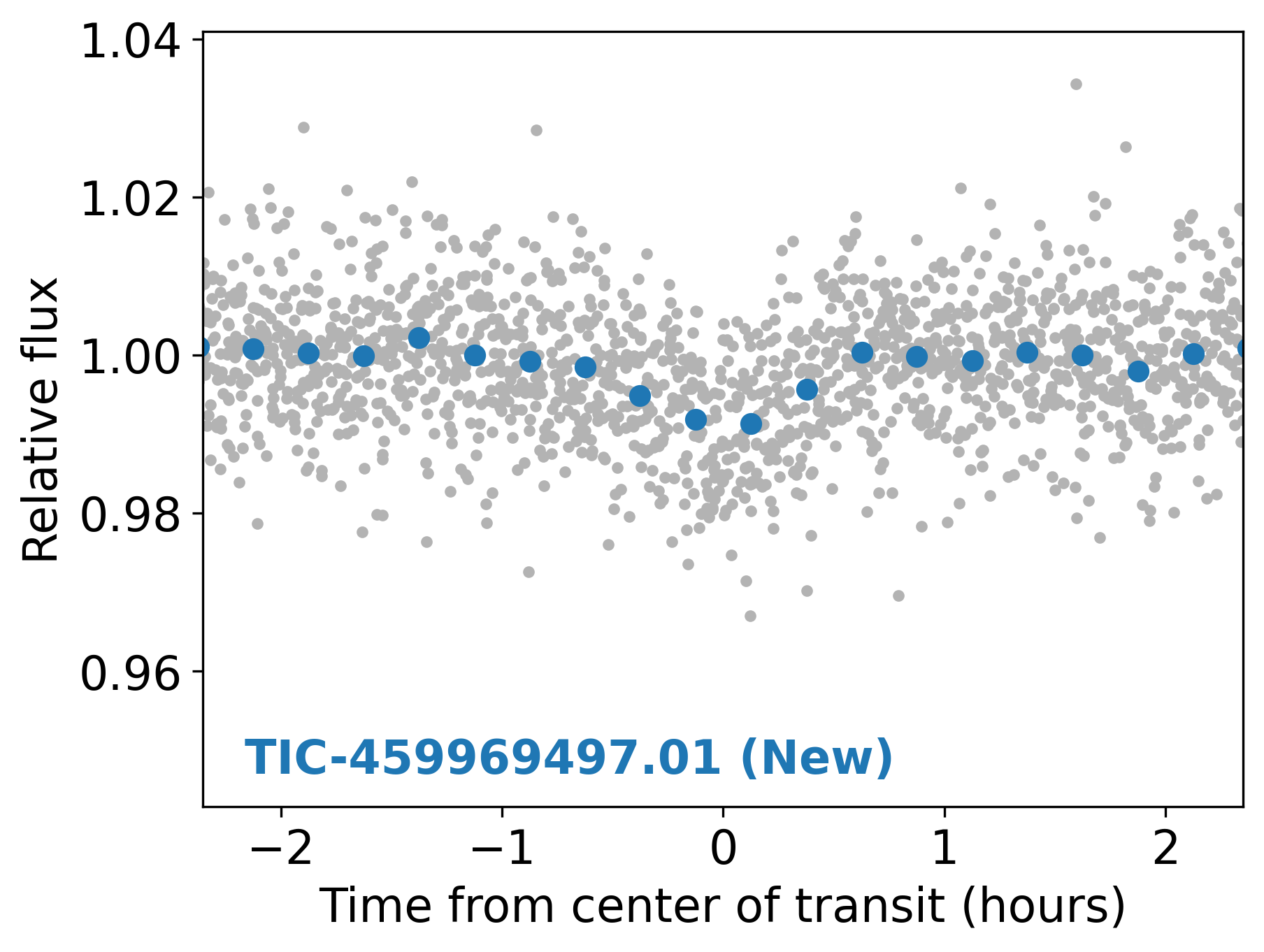}
    \caption{Continued from previous page.}
\end{figure*}

\section{Discussion}

\subsection{Comparison to Other TESS Vetting Tools}

\texttt{LEO-Vetter} is most similar to the publicly available packages \texttt{DAVE} \citep{Kostov2019}, \texttt{EDI-Vetter} \citep{Zink2020}, and \texttt{TEC} \citep{Burke2019}. All four tools were modelled after the Kepler Robovetter and compute metrics that are compared to pass-fail thresholds to reject false positives. While \texttt{DAVE} and \texttt{EDI-Vetter} were originally developed for K2 planet vetting, they have also been used for TESS \cite[e.g.,][]{Boley2021, Feliz2021, Cacciapuoti2022}. \texttt{LEO-Vetter} and \texttt{DAVE} include pixel-level vetting, while \texttt{EDI-Vetter} and \texttt{TEC} do not; however, \texttt{EDI-Vetter} incorporates a flux contamination check to reject blended EBs, and \texttt{TEC} can use centroid results from the SPOC pipeline if such information is already computed.

A major difference between \texttt{LEO-Vetter} and \texttt{DAVE}, \texttt{EDI-Vetter}, and \texttt{TEC} is that the latter three packages do not take into account flux uncertainties. TESS light curves (unlike Kepler and K2) often involve mixes of cadences, and different integration times are associated with different levels of noise (scaled by the square root of the number of data points over a given timescale). By not taking into account flux uncertainties, vetting tests will be biased towards light curve segments with shorter cadences and thus more data points. While users may force TESS light curves to have constant cadences by re-binning shorter cadences to longer cadences, this action results in a loss of information and distinguishing power.

Meanwhile, \texttt{Astronet-Triage-v2} (and its older version \texttt{Astronet-Triage}) is the only other publicly available TESS-specific vetting tool \citep{Yu2019, Tey2023}. Unlike \texttt{LEO-Vetter}, \texttt{Astronet-Triage-v2} is a machine learning classifier using neural networks. \texttt{Astronet-Triage-v2} is able to reject false alarms and contact binaries, but is not able to reject other astrophysical false positives such as non-contact eclipsing binaries and off-target eclipsing objects. Significant automated or manual efforts are still required to vet signals that pass \texttt{Astronet-Triage-v2} \cite[e.g.,][]{Kunimoto2022}.

\subsection{Application to Large-Scale Searches and Demographics}

The first major application of \texttt{LEO-Vetter} is to large planet detection efforts with TESS 20-sec data, 2-min data, FFI data, or any combination of these cadence types. As discussed in Section \ref{sec:leo}, the optimal choice of thresholds will depend on the unique dataset, including the source of the light curves and the stellar sample, as well as the desired balance between completeness and reliability. Users interested in using \texttt{LEO-Vetter} to simply reduce a large number of TCEs down to a manageable list of PCs for further vetting (as is typical for large-scale planet searches) can use our suggested thresholds as a starting point, and then hand-tune them as needed.

For an example use case beyond M dwarfs, we re-did our analysis on the QLP Sector 1 -- 70 light curves for $300,000$ FGK dwarf stars ($R_{\star} < 1.4~R_{\odot}$, $3900 < T_{\mathrm{eff}} < 7300$ K, $\log{g} > 3.5$) brighter than $T = 13$ mag, with 100,000 randomly selected for each spectral class. We searched the original, injected, and scrambled forms of all light curves (resulting in 30,162 observed, 81,287 injected, and 25,570 scrambled TCEs) and computed flux-level vetting metrics following the same procedure as for the M dwarfs. When using the exact same thresholds as for the M dwarfs, we find slightly lower overall completeness (90.86\%) and effectiveness (98.88\%). The overall FGK false alarm reliability is therefore also lower (88.19\%), but still very high for $\mathrm{SNR} > 12$ (98.64\%). We conclude that the thresholds optimized for M dwarfs are also applicable to FGK dwarfs.

\texttt{LEO-Vetter} was also designed to support TESS exoplanet demographic studies, given that statistically robust studies require careful characterization of both catalog completeness and reliability. Completeness must be characterized through both search completeness (the completeness of the planet search process, such as a BLS algorithm for identifying TCEs in light curves) and vetting completeness (the completeness of the vetting algorithm such as \texttt{LEO-Vetter} for identifying PCs from TCEs). Reliability must be characterized against both false alarms and astrophysical false positives. As demonstrated in \S\ref{sec:flux}, \texttt{LEO-Vetter} can be used to characterize vetting completeness and false alarm reliability on a population level. As demonstrated in \S\ref{sec:reliability}, statistical validation packages such as \texttt{TRICERATOPS} can further characterize reliability against astrophysical false positives for individual targets. 

For demographic applications, users can quantitatively optimize thresholds for their unique science purposes because they will have completeness and reliability data products, namely injected and scrambled datasets. We have provided an example of how to run a \texttt{differential\_evolution} algorithm to optimize thresholds given observed, injected, and scrambled metrics in the \texttt{LEO-Vetter} GitHub repository.

Finally, \texttt{LEO-Vetter} has applications to large-scale searches beyond exoplanet science. Because it was designed around tests against false alarms and false positives separately, \texttt{LEO-Vetter} can be used for large EB detection efforts.

\subsection{Future Improvements}\label{sec:future}

\paragraph{Support for other missions} Currently, the transit model fits adopt limb-darkening parameters appropriate only for the TESS bandpass, and pixel-level vetting has only been implemented for TESS FFIs. However, TESS has observing overlap with other space missions such as Kepler and K2, and will have overlap with ESA's PLATO and NASA's Nancy Grace Roman missions. Planet searches and vetting procedures may be applied to data combined from multiple missions, motivating extended functionality to \texttt{LEO-Vetter}.

\paragraph{Supersampling of models} The fluxes associated with each light curve model (straight line, trapezoid, transit, and sine wave models) are computed as though each exposure is instantaneous. An improvement would be to calculate the average value of a given model over an entire exposure using supersampling. When vetting using light curves with mixes of cadences (including those from other missions), the unique exposure times of different light curve segments will need to be considered.

\paragraph{Trapezoid or transit model-based SNR} As described in \S\ref{sec:SES_MES}, \texttt{LEO-Vetter} estimates the SNR of the TCE, an SES time series, and a MES time series, all assuming a box-shaped transit. We chose a box shape because it enables the least computationally expensive estimate of SNR, and the computation of the SES and MES time series is the most expensive part of flux-level vetting. However, a box gives only a rough estimate of SNR that could be improved by using a different shape such as a trapezoid \cite[e.g.,][]{Kipping} or transit model \cite[e.g.,][]{Coughlin2017}.

\paragraph{Combinations of Metrics} Thresholds that take into account correlations between metrics could potentially have stronger distinguishing power than the individual pass-fail thresholds used in this work.

\paragraph{Expansion to Multi-Planet Searches} We demonstrated the vetter's performance based a search of M dwarfs which resulted in only one TCE per star. As mentioned in \S\ref{sec:leo}, \texttt{LEO-Vetter} can also be used for multi-planet searches, but it will treat each TCE independently, meaning that it is currently up to the user to resolve cases such as primary and secondary eclipses being recovered as multi-planet TCEs. Tests against these types of false alarms (e.g., by ephemeris-matching TCEs around the same star) could be added to the vetter in the future. Separately, a future exploration could be to quantify changes in \texttt{LEO-Vetter}'s completeness when vetting multi-planet TCEs. \citet{Zink2019} showed that the search completeness of the Kepler pipeline dropped when recovering additional planets in the same system due to data gaps created by the iterative multi-planet search process, and a similar consequence may be apparent in the vetting process.

\paragraph{Applicability to Transit-Timing Variations} \texttt{LEO-Vetter} assumes that the TCEs are strictly periodic, i.e., they do not feature TTVs. We could explore the consequence of this assumption in future work as TESS TTV catalogs become available.

\section{Conclusions}

We have presented \texttt{LEO-Vetter}, a publicly available Robovetter-inspired tool for fully automated vetting of TESS planet candidates. The vetter employs thirteen flux-level tests against false alarms like stellar variability and instrumental systematics, four flux-level tests against astrophysical false positives like eclipsing binary stars, and pixel-level difference image analysis for identifying off-target signals. Similar to the Kepler Robovetter, \texttt{LEO-Vetter} applies pass-fail thresholds to metrics calculated by these tests to label possible transit signals as either planet candidates (PCs), false alarms (FAs), or false astrophysical positives (FPs). The default pass-fail thresholds presented in this work are suitable for planet searches around FGKM dwarf stars. Users can customize these thresholds for their own science purposes.

By automatically reducing large numbers of possible transit signals down to a manageable list of planet candidates, \texttt{LEO-Vetter} enables analyses that would otherwise be impractical to perform due to time constraints or computational expense. \texttt{LEO-Vetter} was also explicitly designed with statistically robust occurrence rate calculations in mind. As a uniform and fully automated vetting tool, \texttt{LEO-Vetter} enables characterization of both vetting completeness and reliability against false alarms, and produces planet catalogs that can be further characterized for reliability against astrophysical false positives using combinations of follow-up observations and statistical validation packages such as \texttt{TRICERATOPS}.

Here, we demonstrated the vetter's ability to reduce $\sim20,000$ possible transit signals associated with $\sim200,000$ M dwarfs down to a catalog of 172 PCs with both high completeness ($91\%$) and high reliability against false alarms ($97\%$). For each PC, we performed transit modeling, estimated false positive probabilities using \texttt{TRICERATOPS}, identified the most likely source of the transit using a centroid-free Bayesian probability analysis of pixel-level data, and estimated Robovetter-like disposition scores. Following manual inspection, we identified 13 promising new planet candidates that are neither known TOIs nor CTOIs, and we have submitted them as CTOIs to ExoFOP.

\section{Acknowledgments}
% TESS and SPOC
We thank the referee for constructive comments which improved the paper. Funding for the TESS mission is provided by NASA's Science Mission Directorate. We acknowledge the use of public TESS data from pipelines at the TESS Science Office and at the TESS Science Processing Operations Center. This paper includes data collected by the TESS mission that are publicly available from the Mikulski Archive for Space Telescopes (MAST). Quick-Look Pipeline light curves are available as High-Level Science Products (HLSP) on MAST \citep{QLPHLSP}.

%SPOC
Resources supporting this work were provided by the NASA High-End Computing (HEC) Program through the NASA Advanced Supercomputing (NAS) Division at Ames Research Center for the production of the SPOC data products.

% ExoFOP
This research has made use of the Exoplanet Follow-up Observation Program (ExoFOP; DOI: 10.26134/ExoFOP5) website, which is operated by the California Institute of Technology, under contract with the National Aeronautics and Space Administration under the Exoplanet Exploration Program.

% Individuals
MK acknowledges the support of the Natural Sciences and Engineering Research Council of Canada (NSERC), RGPIN-2024-06452. Cette recherche a été financée par le Conseil de recherches en sciences naturelles et en génie du Canada (CRSNG), RGPIN-2024-06452. MK also acknowledges support by the Juan Carlos Torres postdoctoral fellowship from the MIT Kavli Institute for Astrophysics and Space Research. 

TD acknowledges support from the McDonnell Center for the Space Sciences at Washington University in St. Louis.

JFR acknowledges financial support from NSERC Discovery, teaching relief from Canada Research Chairs program and computing facilities from Digital Alliance Canada.

YNEE acknowledges support from a Science and Technology Facilities Council (STFC) studentship, grant number ST/Y509693/1.

\vspace{5mm}
\facilities{TESS}
\software{\texttt{batman} \citep{Kreidberg2015}, \texttt{cuvarbase} \citep{Hoffman2022}, \texttt{lmfit} \citep{LMFIT}, \texttt{matplotlib} \citep{Hunter2007}, \texttt{numpy} \citep{Harris2020}, \texttt{pandas} \citep{reback2020pandas, mckinney-proc-scipy-2010},  \texttt{scipy} \citep{Virtanen2020}, \texttt{tess-point} \citep{tesspoint},  \citep{Giacalone2021}, \texttt{wotan} \citep{Hippke2019}}

\bibliography{sample7}{}
\bibliographystyle{aasjournal}

\end{document}